\documentclass[usenatbib,a4paper]{mn2e}
\usepackage{graphicx}
\usepackage{epstopdf}  
\usepackage{amssymb}  
\usepackage{booktabs,caption,fixltx2e}  
\usepackage[flushleft]{threeparttable}  

\title[The BNE and its application to asteroids]{The Brazil-nut effect and its application to asteroids}

\author[Matsumura et al.]
{Soko Matsumura,$^{1,2}$\thanks{E-mail: s.matsumura@dundee.ac.uk}
Derek C. Richardson,$^{2}$
Patrick Michel,$^{3}$ 
\newauthor
Stephen R. Schwartz$^{3,2}$ and
Ronald-Louis Ballouz$^{2}$ \\
$^{1}$School of Engineering, Physics, and Mathematics, University of Dundee, Scotland DD1 4HN\\
$^{2}$Department of Astronomy, University of Maryland, College Park, MD 20741\\
$^{3}$Lagrange Laboratory, University of Nice-Sophia Antipolis, CNRS, Observatoire de la C\^{o}te d'Azur, France
}

\pagerange{\pageref{firstpage}--\pageref{lastpage}} \pubyear{2014}

\begin{document}

\label{firstpage}

\maketitle

\begin{abstract}
Out of the handful of asteroids that have been imaged, some have 
distributions of blocks that are not easily explained.
In this paper, we investigate the possibility that seismic shaking leads to the size sorting of particles in asteroids.
In particular, we focus on the so-called Brazil Nut Effect (BNE) that separates 
large particles from small ones under vibrations.
We study the BNE over a wide range of parameters by using the {\it N}-body code PKDGRAV, 
and find that the effect is largely insensitive to the coefficients of restitution, 
but sensitive to friction constants and oscillation speeds.
Agreeing with the previous results, we find that convection drives the BNE, where the intruder rises to the top of the particle bed. 
For the wide-cylinder case, we also observe a ``whale'' effect, where the intruder follows the convective current and does not stay at the surface.
We show that the non-dimensional critical conditions for the BNE 
agree well with previous studies.
We also show that the BNE is scalable for low-gravity environments 
and that the rise speed of an intruder is proportional to the square root of the gravitational acceleration. 
Finally, we apply the critical conditions to observed asteroids, 
and find that the critical oscillation speeds are 
comparable to the seismic oscillation speeds that are expected from non-destructive impacts.
\end{abstract}

%
\section{Introduction}\label{intro}
Asteroids are small rocky bodies left over from the planet formation era.
Although their internal structure is not well constrained, the images of a handful of 
asteroids have provided us with the clues to the origins of their surface features 
\cite[e.g.,][]{Asphaug2002ASTE,Chapman2004AREPS}.  
Perhaps the most prominent feature is craters that indicate the history of impacts by other objects.
However, there are other surface features that are more subtle and need explanations.

For example, \shortcite{Asphaug2001LPI} pointed out that the blocks on (433) Eros do 
not obey any obvious dynamical distribution such as 
association with potential source craters or 
sweep-up correlated with the asteroid rotation.
These blocks are likely ejecta fragments, but most of them are not seen with pits that would indicate 
collisions at meters per second into the low-gravity regolith.
They proposed that size sorting in asteroid regolith could explain such a distribution of blocks. 

Another potential example of this size sorting is (25143) Itokawa, 
where smooth and rugged regions are observed.
\shortcite{Miyamoto2007Sci} proposed that the substantial vibrations and 
the resulting granular convection could explain such a difference in distributions of particles.  
Besides the abundance of boulders, Itokawa 
is also characterized by a lack of major craters \cite[][]{Fujiwara2006Sci}. 
Since the total volume of boulders on Itokawa is larger than the available volume 
in the identified craters, all the boulders cannot be from these craters.
This may further support the size-sorting scenario on Itokawa.
Alternatively, the absence of large fresh craters might be explained 
if the most recent impact erased its own crater \cite[e.g.,][]{Michel2009Icar}.
Furthermore, the boulders on Itokawa could also be explained by assuming that Itokawa 
was made by the reaccumulation of fragments from a catastrophically 
disrupted parent body \cite[][]{Michel2013AA}. 
In this paper, we explore whether the size sorting due to global oscillations is possible on 
asteroids such as Eros and Itokawa.

\shortcite{Asphaug2001LPI} identified two phenomena in granular systems that might have direct 
relevance to asteroids.
One of them is the inelastic segregation that could form regions of granular solid via 
the aggregation of inelastically colliding objects. 
The other is the size segregation via the so-called Brazil Nut Effect (BNE) 
that describes the rise of a particle embedded in an oscillating system of 
smaller particles in a confined environment \cite[e.g.,][]{Rosato1987PhRvL}.
We focus on this latter effect in this paper.
The BNE has been studied in many different setups, but rarely in the low-gravity 
environments that are relevant to asteroids. 
There are some recent exceptions, such as \shortcite{Tancredi2012MNRAS} and \shortcite{Guttler2013PhRvE}, 
and we compare our results with theirs in Sections~\ref{results_Tancredi12} and 
\ref{results_lowg}.

There are two distinct models of the BNE --- one of them is the intruder model, 
where the number of large particles is small compared to the number of smaller ones, and 
the other is the binary mixture model, where both small and large particles occupy 
comparable volumes. These models behave differently under vibrations, partly because 
interactions between large particles become significant for the latter 
\cite[e.g.,][]{Sanders2004PhRvL}.
The internal structure of asteroids is not well known, and has so far been only guessed at 
through theoretical and numerical works \cite[e.g.,][]{Asphaug2002ASTE,Richardson2002ASTE}.
Due to this lack of knowledge, we assume in this paper that the intruder model is appropriate 
for asteroids.

The behavior of the intruder model differs depending on the oscillation speeds 
\cite[e.g.,][]{Kudrolli2004RPPh}.
When the oscillation is weak, the particles are in the dense limit, where 
the contacts between particles last for a long time.
As the oscillation speed increases, the system can become vibro-fluidized, 
where the particles behave like a fluid and their interactions can be treated 
as binary collisions.
In the vibro-fluidized limit, the intruder's behavior depends on the size ratio 
and the density ratio of the constituent particles.
The intruder rises to the surface when its density is lower than that of the surrounding 
smaller particles, and sinks when its density is larger 
\cite[e.g.,][]{Ohtsuki1995JPSJ,Hsiau1997PT,Shishodia2001PhRvL,Breu2003PhRvL,Huerta2004PhRvL}.
In the dense limit, the intruder rises independent of the density ratio 
\cite[e.g.,][]{Liffman2001GM,Huerta2004PhRvL}. 
All of our simulations are likely to be in the dense limit most of the time, because 
most particles have a number of neighboring particles in contact.
As we describe in the next section, we use a version of the {\it N}-body code PKDGRAV 
that can handle systems with long-lasting contacts. 

In this paper, we investigate the Brazil Nut Effect as a potential mechanism to sort 
particles in asteroids and to explain distributions of boulders on asteroids' surfaces.
The efficiency of the BNE depends on many different parameters; furthermore, those corresponding 
parameters in asteroids are unknown.
Therefore, we first investigate the efficiency of the BNE through 
a wide range of initial conditions and then apply the model to the low-gravity 
environments that are suitable for asteroids.   
In Section~\ref{method}, we introduce our numerical methods and the choice of initial conditions. 
In Section~\ref{results}, we study the dependence of the BNE on various parameters including 
the coefficients of restitution, the friction constants, the oscillation speeds, 
and the depth of particle beds (Sections~\ref{results_eps}--\ref{results_vel}). 
We also compare our models with previous studies and find good agreements 
(Sections~\ref{results_vel} and \ref{results_Tancredi12}).
The BNE in low-gravity environments is investigated in Section~\ref{results_lowg}, and 
we apply the critical conditions to observed asteroids.
Finally in Section~\ref{summary}, we discuss and summarize our study.
%
%
%
%
\section{Method}\label{method} 
In this section, we introduce our
numerical code PKDGRAV (Section~\ref{method_pkdgrav}) as well as the
initial conditions used for our simulations (Section~\ref{method_ic}).
\subsection{Numerical Method: PKDGRAV}\label{method_pkdgrav}
PKDGRAV is a parallel $N$-body gravity tree code \cite[][]{Stadel2001PhDT} 
adapted for particle collisions 
\cite[][]{Richardson2000Icar,Richardson2009PSS,Richardson2011Icar}.  
Originally, collisions in PKDGRAV were treated as
idealized single-point-of-contact impacts between rigid spheres.  
However, such an instantaneous collision assumption is not appropriate 
for a dense system like a particle bed, where particle contacts can last many 
timesteps.
Recently, \shortcite{Schwartz2012GM} added a soft-sphere option to PKDGRAV that handles 
long-lasting contacts with reaction forces dependent on the degree of overlap 
(a proxy for surface deformation) and contact history. 
We use this version of the code for our study of the BNE.
The code uses a 2nd-order leapfrog integrator, with accelerations due to gravity and contact
forces recomputed each step.  Various types of user-definable
confining walls are available that can be combined to provide complex
boundary conditions for the simulations.  
For example, we use an infinitely tall cylinder and box as our container, 
as described in the following subsection.
The code also includes an optional variable gravity field based on a user-specified set of
rules.
This allows us to change the magnitude and a direction of gravity in the simulation 
(see Section~\ref{results_lowg} for details). 

The spring/dash-pot model used in PKDGRAV's soft-sphere
implementation is described fully in \shortcite{Schwartz2012GM}.  Briefly,
a (spherical) particle overlapping with a neighbor or confining wall
feels a reaction force in the normal and tangential directions
determined by spring constants ($k_n$, $k_t$), with optional damping
and effects that impose static, rolling, and/or twisting friction.
The damping parameters ($C_n$, $C_t$) are related to the conventional
normal and tangential coefficients of restitution used in hard-sphere
implementations, $\epsilon_n$ and $\epsilon_t$.  
The static, rolling, and twisting friction components are parameterized by
dimensionless coefficients $\mu_s$, $\mu_r$, and $\mu_t$,
respectively.  Plausible values for these various parameters are
obtained through comparison with laboratory experiments.  
Careful consideration of the soft-sphere parameters is needed to ensure
internal consistency, particularly with the choice of $k_n$, $k_t$,
and timestep --- a separate code is provided to assist the user with
configuring these parameters correctly.  
For most of our simulations, $k_n=2.261947\times10^9\,{\rm g \, s^{-2}}$ and 
timestep $1.851201\times10^{-6}\,$s are adopted, 
which permits a maximum particle speed of $200\,{\rm cm \, s^{-1}}$ with 
no more than 1 \% overlap between spheres. 
As a conservative estimate, we consider a maximum particle speed that 
is larger than the maximum oscillation speed of our default 
case of $93.9\,{\rm cm \, s^{-1}}$. 
For other parameters, $\mu_t$ is set to zero all the time for simplicity, 
while $\epsilon_n$, $\epsilon_t$, $\mu_s$ and $\mu_r$ are varied over wide ranges.
We will investigate the effect of $\mu_t$ in future work.

The numerical approach has been validated through comparison with laboratory experiments; 
e.g., \shortcite{Schwartz2012GM} demonstrated that PKDGRAV correctly
reproduces experiments of granular flow through cylindrical hoppers,
specifically the flow rate as a function of aperture size, and 
\shortcite{Schwartz2013Icar} demonstrated successful simulation of
laboratory impact experiments into sintered glass beads using a
cohesion model coupled with the soft-sphere code in PKDGRAV.
Most recently, \shortcite{Schwartz2014inprep} applied the code to low-speed impacts into 
regolith in order to test asteroid sampling mechanism design.
\subsection{Initial Conditions}\label{method_ic}
Unless noted otherwise, we use the same particle distributions in
an infinitely tall cylinder with a diameter of $10\,$cm 
as the initial condition for all our simulations. 
In our intruder model, there are 1800 small particles and 1 larger particle, where the
diameters of small and large particles are $d_s=1\,$cm and $d_l=3\,$cm,
respectively. Both types of particles are assumed to have the same mass density of
$2.7\,{\rm g \, cm^{-3}}$, which corresponds to the density of aluminium. For our default
simulation, we adopt the coefficients of restitution
$\epsilon_n=\epsilon_t=0.5$, static friction $\mu_s=0.7$, and rolling
friction $\mu_r=0.1$.
\footnote{We note that $\epsilon_t$ used here is not the true tangential coefficient of restitution which is 
difficult to specify in soft-sphere simulations \cite[see][]{Schwartz2012GM}.
Still, $\epsilon_t$ has a one-to-one mapping to a dimensionless quantity $C_t$ 
as mentioned in Section~\ref{method_pkdgrav} and is defined as 
$C_t \equiv -2\ln \epsilon_t\sqrt{k_t\mu/(\pi^2+(\ln\epsilon_t)^2)}$.} 
The choices of parameters are rather arbitrary.
However, as we see below, the BNE is relatively insensitive to the exact choice of the coefficients of restitution. 
The BNE is sensitive to the choice of the friction constants, so 
we choose the values that result in the occurrence of BNE as default values.
In terms of real materials, these constants nearly correspond to the oxidized Al 
that was used in the experiments of \shortcite{Clement1992PhRvL}, 
where they studied a vibrationally excited pile of beads and observed spontaneous heaping and convection. 
Besides particle-particle interactions that are described above, PKDGRAV also handles 
particle-wall interactions.  
We use the same coefficients of restitution and friction constants as the particles for 
all of the cylinder walls.

\begin{figure}
\includegraphics[width=84mm]{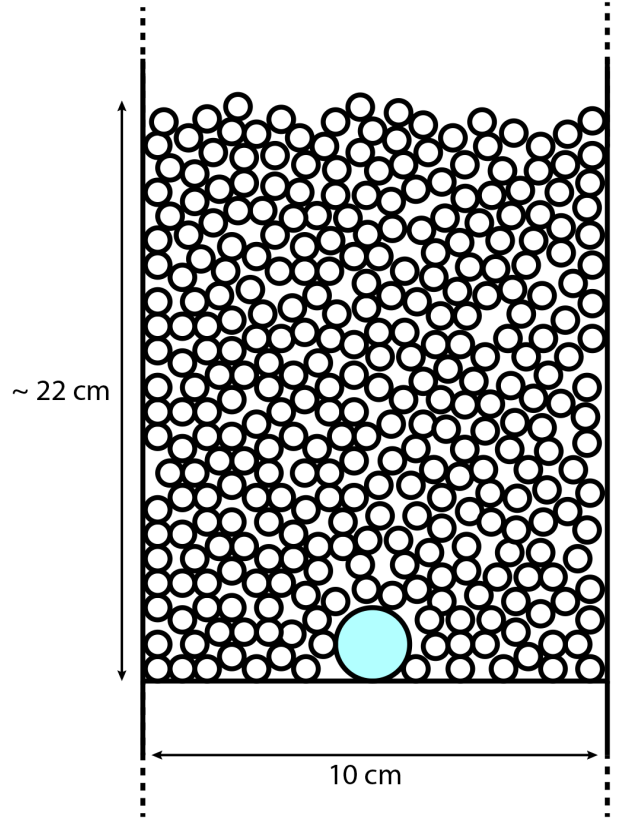}
\caption{Schematic diagram of a cross-section of the experiment after an infinitely 
long cylindrical container (with a diameter of $10\,$cm) is filled up with small particles 
(up to $\sim 22\,$cm from the origin), but before shaking begins. 
The large cyan particle (the intruder) is initially located at the floor.
\label{fig_ron}}
\end{figure}

The simulations are divided into two stages --- the filling of the
cylinder with particles, and the oscillation of the filled cylinder. In
the first stage, the large particle is initially placed at the floor 
of the cylinder (whose center is the origin of the coordinate), 
while small particles are suspended in the air, $10-130\,$cm above the bottom panel. 
The free-fall simulation is done under Earth gravity $g=980\,{\rm cm \ s^{-2}}$, 
and the cylinder is filled with small particles in $\sim0.5\,$s. 
For the low gravity simulations in Section~\ref{results_lowg}, 
this stage is done under the corresponding gravitational environment. 
Under the influence of Earth gravity, particles fill to a height of just under $22\,$cm.
The schematic figure of our system at the end of the first stage is shown in Figure~\ref{fig_ron}.
In the second stage, the entire cylinder is vertically shaken in a sinusoidal manner
with a given amplitude $A$ and frequency $\omega$ as $z=A\sin(\omega t)$, 
where $z$ is the height of the base of the cylinder relative to an arbitrary zero point. 
The default amplitude and frequency are $A=d_s=1\,$cm, and
$\omega=3\sqrt{a_g/A}=93.9\,{\rm rad \, s^{-1}}$. 
Here, $a_g$ is a gravitational acceleration, and $a_g=g$ is assumed for most cases 
except for Section~\ref{results_lowg}.
Thus, the maximum speed for the
default shake is $v_{max}=\omega A=93.9\,{\rm cm \, s^{-1}}$.
Most of our simulations are done on a single core of a modern CPU such 
as Intel Xeon and AMD Opteron. 
For the simulations in this paper, the runtime varied from $\sim 10\,$hours 
to $\sim10\,$days for 150 cycles, 
depending on the number of particles as well as the choice of parameters such as oscillation frequencies. 
In general, low oscillation frequency simulations take longer than high frequency ones, 
because a longer time is required to complete 150 cycles of oscillation.
%
%
\section{Results}\label{results} 
In this section, we present the results 
of our simulations. First, we vary several parameters and investigate
how they affect the BNE. 
\subsection{Dependence on Coefficients of Restitution}\label{results_eps} 
In this subsection, we study the 
dependence of BNE on normal and tangential coefficients of restitution
($\epsilon_n$ and $\epsilon_t$, respectively). For all the simulations
in this subsection, we assume that the static and rolling friction
constants are $\mu_s=0.7$ and $\mu_r=0.1$, respectively, and that the
oscillation amplitude and frequency are $A=d_s=1\,$cm and
$\omega=3\sqrt{g/d_s}=93.9\,{\rm rad \, s^{-1}}$, respectively.

\begin{figure}
\includegraphics[width=84mm]{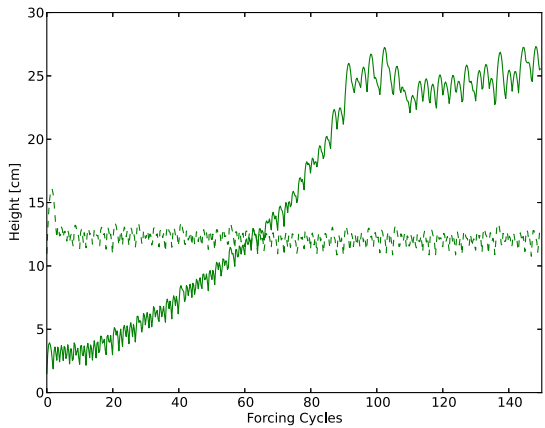}
\caption{Height evolution of the intruder (solid line) compared with
evolution of the median height of the small particles (dashed line). Here,
$\epsilon_n=\epsilon_t=0.5$, $\mu_s=0.7$ and $\mu_r=0.1$ are assumed.
The intruder rises from the bottom of the cylinder to the surface of the
particle bed in $\sim90$ cycles. \label{fig_epsn0.5_epst0.5}}
\end{figure}

The height evolution during the oscillation simulation for our default
case is shown in Figure~\ref{fig_epsn0.5_epst0.5}. The intruder's height
is compared to the median height of small particles $\bar{z}_s$.
After $\sim90$ cycles of the oscillations, the intruder rises to the top of the
particle bed. 
Since the height of the particle beds is constantly changing due to oscillations, 
we define that the intruder arrives at the surface when $z_l>2 \bar{z}_s$.
To estimate the variation in our oscillation simulations, we generate 10 
independent particle distributions with 1800 small particles and 1 intruder, 
and perform oscillation simulations by using the same parameters for all of them.
For the default parameter set, we find that the average rise cycle (number of 
oscillations needed for the intruder to rise to the top) is 
$\bar{\tau}_{cyc}=92.0$ and the standard deviation is $\sigma=6.7$.

\begin{figure}
\includegraphics[width=84mm]{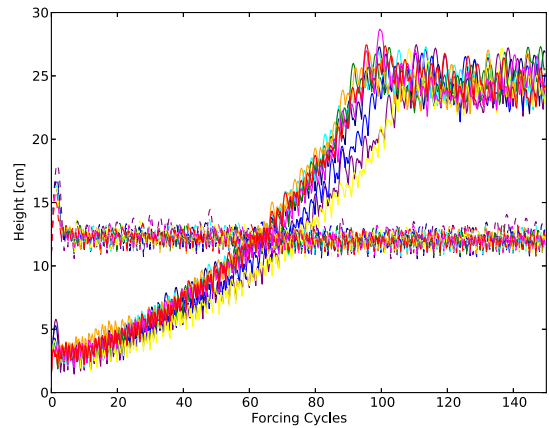}
\caption{Height evolution of the intruder (solid line) for
$\epsilon_n=0.5$ and $\epsilon_t=0.1-0.9$. All of the rise cycles are
consistent with the mean rise cycle determined from the default case within $\sim3\sigma$, 
where $\sigma$ is the default case's standard deviation that is defined in text. 
The line colors of red, magenta, orange, yellow, green, cyan, blue, navy, and
purple correspond to $\epsilon_t=0.1-0.9$ in an increasing order. 
\label{fig_epsn0.5_epst0.1-0.9}} 
\end{figure}

Now we explore the different sets of restitution coefficients.
We change normal and tangential coefficients of restitution from 
$0.1-0.9$ with $\Delta\epsilon=0.1$, and find that the BNE occurs in all of the 81 cases. 
Moreover, we find that the rise time of an intruder seems relatively independent of the 
choices of coefficients of restitution. 
In Figure~\ref{fig_epsn0.5_epst0.1-0.9}, we compare simulations with
$\epsilon_n=0.5$ and $\epsilon_t=0.1-0.9$ as an example. 

\begin{figure*}
\includegraphics[width=0.48\textwidth]{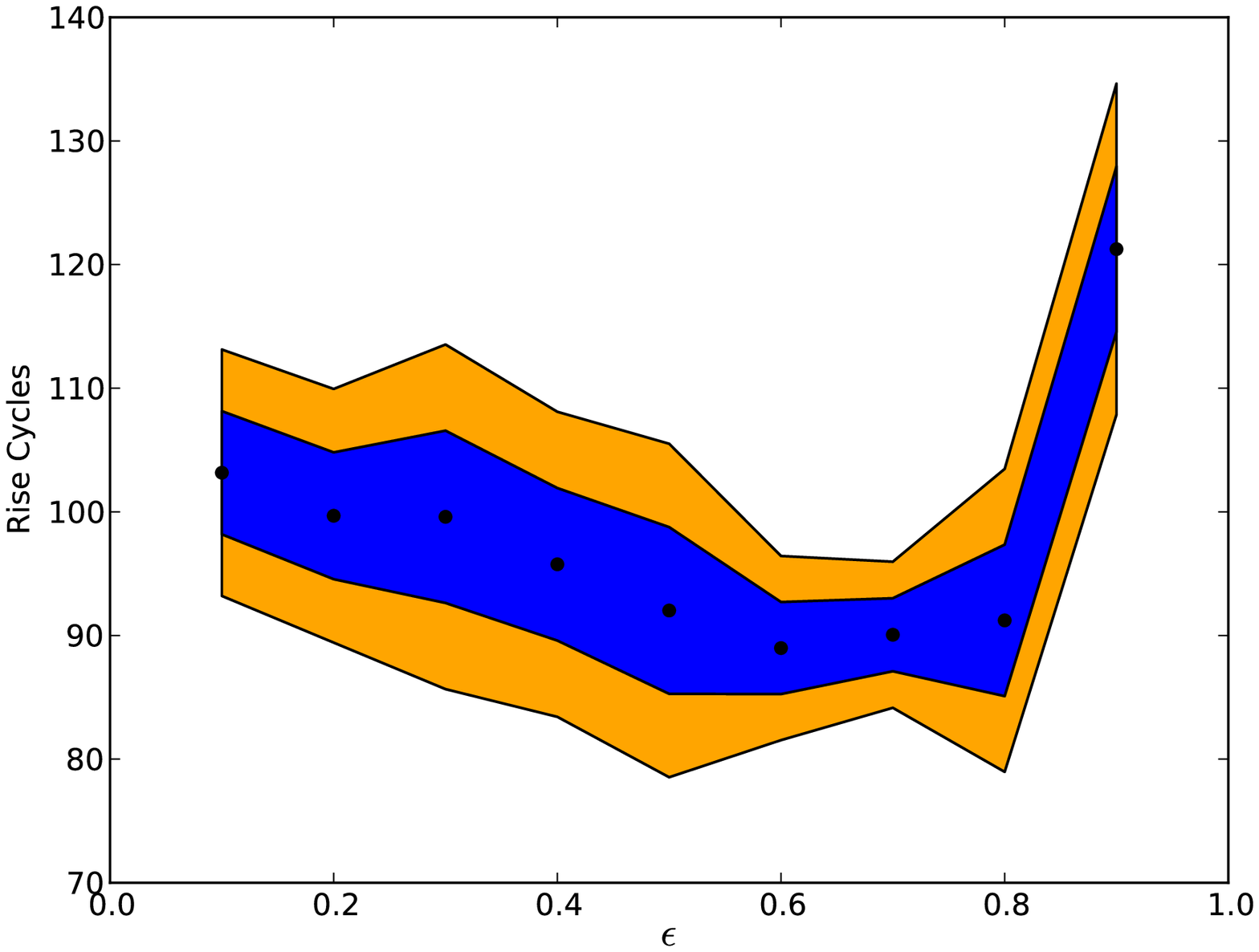}
\includegraphics[width=0.48\textwidth]{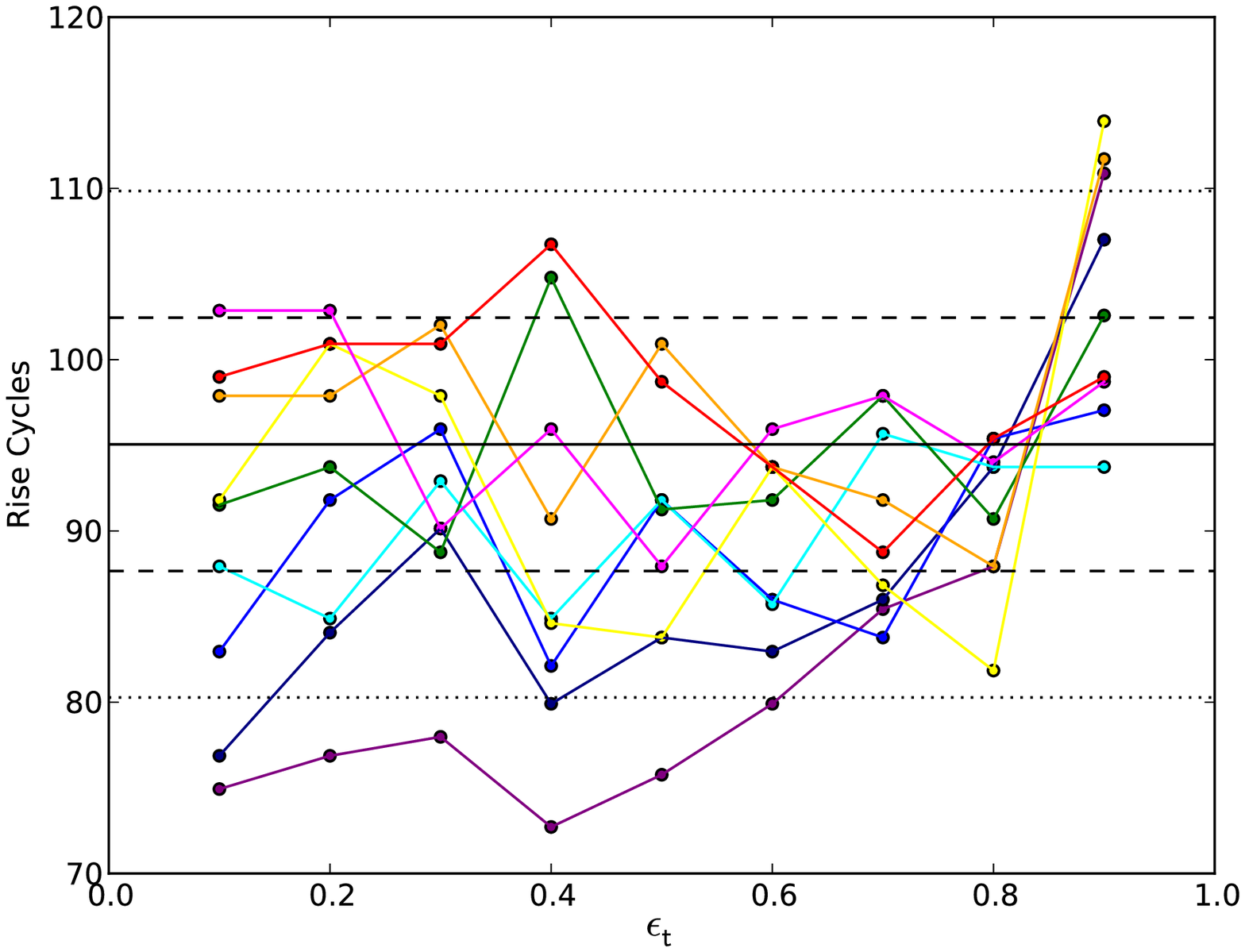}
\caption{Left: The mean rise cycle (circles) and the standard deviation $\sigma$ estimated from 10 simulations each 
for $\epsilon_n=\epsilon_t=0.1-0.9$.  The blue and orange regions correspond to $1\sigma$ and $2\sigma$, respectively.
Except for $\epsilon_n=\epsilon_t=0.9$, all of these sets have variations in the rise cycles that are 
consistent within $2\sigma$.
Right: the rise cycles for simulations with various $\epsilon_n$ and $\epsilon_t$. 
The red, magenta, orange, yellow, green, cyan, blue, navy, and
purple lines correspond to $\epsilon_n=0.1-0.9$ in an increasing order. 
The black solid, dashed, and dotted lines indicate the mean rise cycle, $1\sigma$, and $2\sigma$ estimated from 
80 simulations with $\epsilon_n=\epsilon_t=0.1-0.8$.
\label{fig_sigma_epsn_epst}} 
\end{figure*}

To understand this similarity further, we estimate the variations in oscillation simulations by using 
10 different particle distributions for 9 different combinations of restitution coefficients 
$\epsilon_n=\epsilon_t=0.1-0.9$. 
The mean rise cycle and the standard deviation for each set is compared in the left panel of Figure~\ref{fig_sigma_epsn_epst}.
The mean rise cycle decreases from $\epsilon_n=\epsilon_t=0.1$ to $\epsilon_n=\epsilon_t=0.6$ and then slightly increases 
from $\epsilon_n=\epsilon_t=0.6$ to $\epsilon_n=\epsilon_t=0.8$.  
The rise cycle for $\epsilon_n=\epsilon_t=0.9$ increases very sharply from $\epsilon_n=\epsilon_t=0.8$.
In fact, the variations of the rise cycles are consistent within $2\sigma$ for all the cases except for $\epsilon_n=\epsilon_t=0.9$. 

The right panel of Figure~\ref{fig_sigma_epsn_epst} shows the rise cycle estimated for each combination of $\epsilon_n$ and $\epsilon_t$ 
by using the same particle distribution.
We also plot the mean cycle and the variations estimated from 80 simulations with $\epsilon_n=\epsilon_t=0.1-0.8$. 
The rise cycles of many combinations appear within $1\sigma$ from the mean cycle, and most appear within $2\sigma$.
A clear exception is the $\epsilon_n=0.9$ cases that tend to have shorter rise cycles than the others 
for small values of $\epsilon_t$. 
There is an indication that the rise time might become shorter for $\epsilon_n=0.9$ 
and longer for $\epsilon_t=0.9$. 
It is unclear whether this represents poor sampling or a true trend.
However, we should note that the BNE does not occur when the collisions are perfectly elastic 
either in normal or in tangential direction (i.e., either $\epsilon_n$ or $\epsilon_t$ is 
1.0).
It is possible that systems with bouncier particles behave differently 
from those with less elastic ones, 
because such systems could transit from the dense system to the vibro-fluidized system 
at a lower oscillation frequency.
Since our goal here is to understand the overall trend of the BNE, 
we defer a more detailed investigation on high values of coefficients of restitution 
for future work. 

The coefficients of restitution for asteroid constituents are not well constrained.
Recently, \shortcite{Durda2011Icar} performed head-on collision
experiments between two granite spheres with diameters of 1 m, and
estimated that the coefficient of restitution is
$\epsilon_n=0.83\pm0.06$ for collision speeds up to $\sim1.5\,{\rm m \, s^{-1}}$.
This normal coefficient of restitution value is relatively large, but is still expected to 
lead to a general BNE behavior, independent of the value of $\epsilon_t$. 

In summary, our results indicate that the BNE is largely
independent of the exact choices of coefficients of restitution, except in the most 
elastic cases. 
Thus, for the rest of the paper, we assume $\epsilon_n=\epsilon_t=0.5\,$ 
unless it is noted otherwise.
\subsection{Dependence on Friction}\label{results_mu} 
In this subsection, we study the dependence of BNE on static and rolling
friction constants ($\mu_s$ and $\mu_r$, respectively). For all the
simulations in this subsection, we assume that the normal and tangential
coefficients of restitution are $\epsilon_n=\epsilon_t=0.5$, and that
the oscillation amplitude and frequency are $A=d_s=1\,$cm and
$\omega=3\sqrt{g/d_s}=93.9\,{\rm rad \, s^{-1}}$, respectively.

We change the static friction over $\mu_s=0.0-1.0$ and rolling friction
over $\mu_r=0.0-0.2$. 
We note that, for cohesionless materials, the static friction coefficient is 
related to the angle of 
repose of the loose material by $\tan\phi = \mu_s$.  Thus, $\mu_s=1.0$ corresponds to 
material with a relatively high (45-degree) angle of repose, and sampling from 
$\mu_s=0.0$ to $1.0$ covers a good range of plausible material properties.
The rolling friction does not have as nice a physical correspondence as $\mu_s$, but 
the friction comes in as a torque due to the normal force acting at a contact point of 
two particles \cite[][]{Schwartz2012GM}.

\begin{figure} 
\includegraphics[width=84mm]{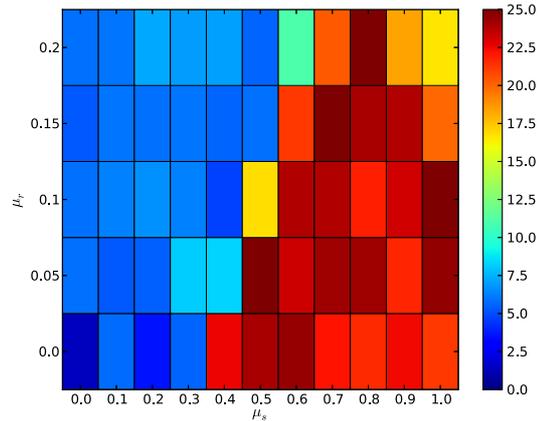}
\caption{Final intruder heights after 150 oscillations as a 
function of $\mu_s$ and $\mu_r$.
Each rectangular region represents a specific choice of $\mu_s$ and $\mu_r$ and 
is color coded by the final intruder height (see color legend to the right of the plot).
For all the simulations, coefficients of restitution are set to
$\epsilon_n=\epsilon_t=0.5\,$. 
There is a sharp transition from no-BNE to BNE regions. 
A threshold static friction is necessary for the BNE to occur, but 
high static or rolling friction diminishes the BNE.
 \label{fig_mus_mur}} 
\end{figure}

\begin{figure*}
\includegraphics[width=0.48\textwidth]{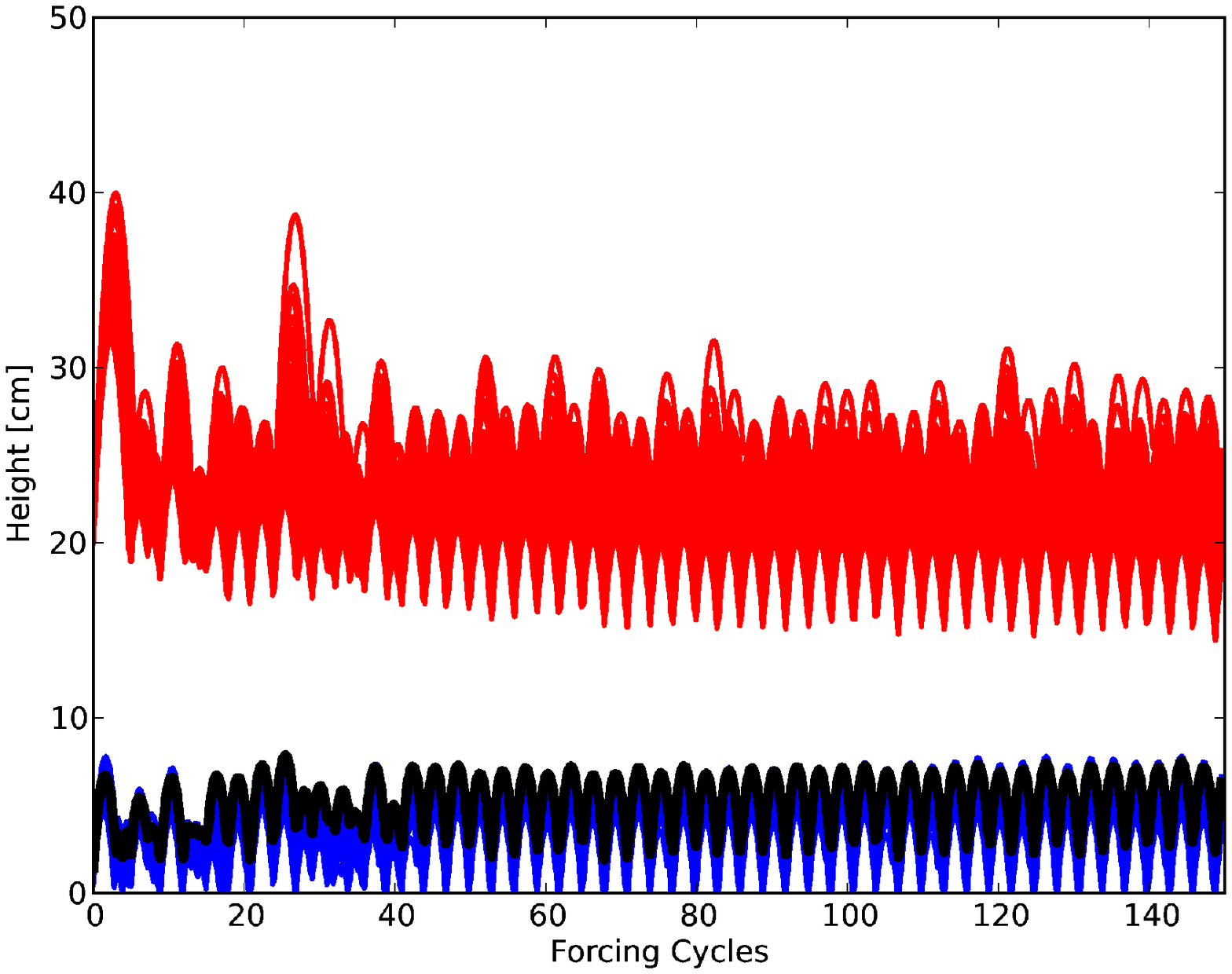}
\includegraphics[width=0.48\textwidth]{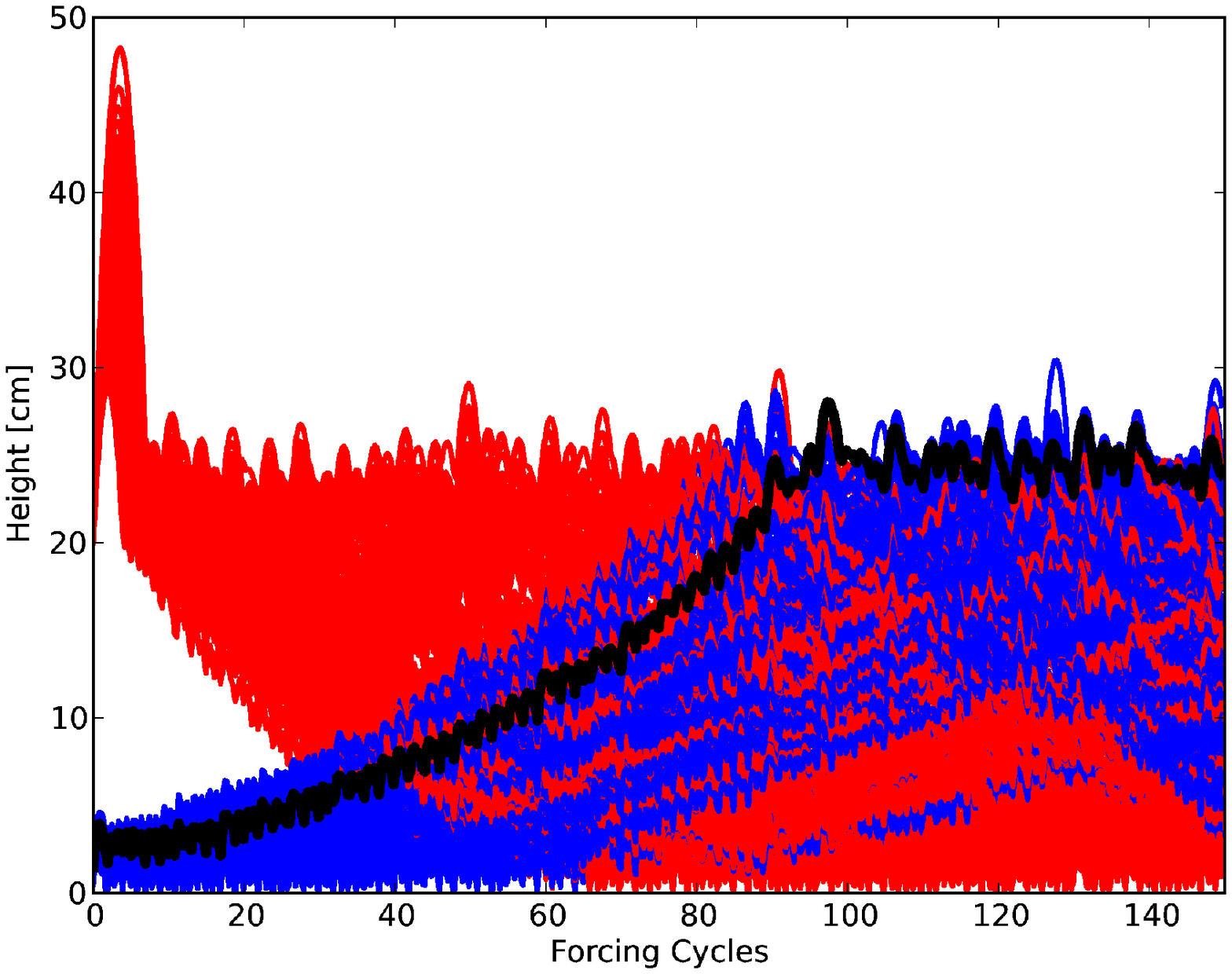}
\caption{Height evolutions of small particles in the
uppermost layer (red) and the lowermost layer (blue) are compared with
the corresponding evolution of the intruder (black). All the parameters are the
same except that $\mu_s=0.3$ and $\mu_s=0.7$ in the left and right
panels, respectively. When the BNE does not occur (left), particle
layers are well-separated throughout the simulation. On the other hand,
when the BNE occurs (right), particle layers are well-mixed due to
convection.  \label{fig_mus0.3_mus0.7}} 
\end{figure*}

In Figure~\ref{fig_mus_mur}, we plot the instantaneous heights
of intruders after 150 oscillations for each simulation. We find that
the efficiency of the BNE depends steeply on the friction constants. The
figure indicates that the BNE requires a high enough $\mu_s\gtrsim 0.5$ and a low
enough $\mu_r\lesssim0.2$.
The difference between non-BNE and BNE regions is illustrated in
Figure~\ref{fig_mus0.3_mus0.7}. By dividing the initial distributions of
particles into 11 layers (i.e., each layer is $\sim2\,$cm thick), we
plot the height evolution of particles which are initially in the
uppermost and lowermost layers, along with that of the intruder
particle. There is little vertical mixing of particles when there is no
BNE (left panel), while particles are well-mixed when the BNE is
observed (right panel). Convection of particles is observed in all
the simulations with the BNE, where particles descend along the wall and 
ascend in the central region. 
More precisely, small particles follow gradual rise and fall 
cycles throughout the entire particle distribution while the intruder rises but 
does not fall. 
Our results agree with many previous works
that have observed convection along with the BNE 
\cite[e.g.,][]{Knight1993PhRvL,Poschel1995EL}. 
Furthermore, the trend seen in Figure~\ref{fig_mus_mur} agrees with the implications of \shortcite{Clement1992PhRvL}.
They experimentally investigated a two-dimensional pile of equal-sized beads under a vertical sinusoidal oscillation, 
and found that the convection is not observed for polished aluminum beads with $\mu_s=0.2$, but 
is observed for oxidized ones with $\mu_s=0.8$. In their experiments, both kinds of beads have a normal restitution coefficient of 0.5, 
which is comparable to our default value.
In our simulations, for $\mu_s=0.2$, the BNE was not observed for any value of $\mu_r$, 
while for $\mu_s=0.8$, the BNE was seen for all the values of $\mu_r$ we tested.

\begin{figure*}
\includegraphics[width=0.48\textwidth]{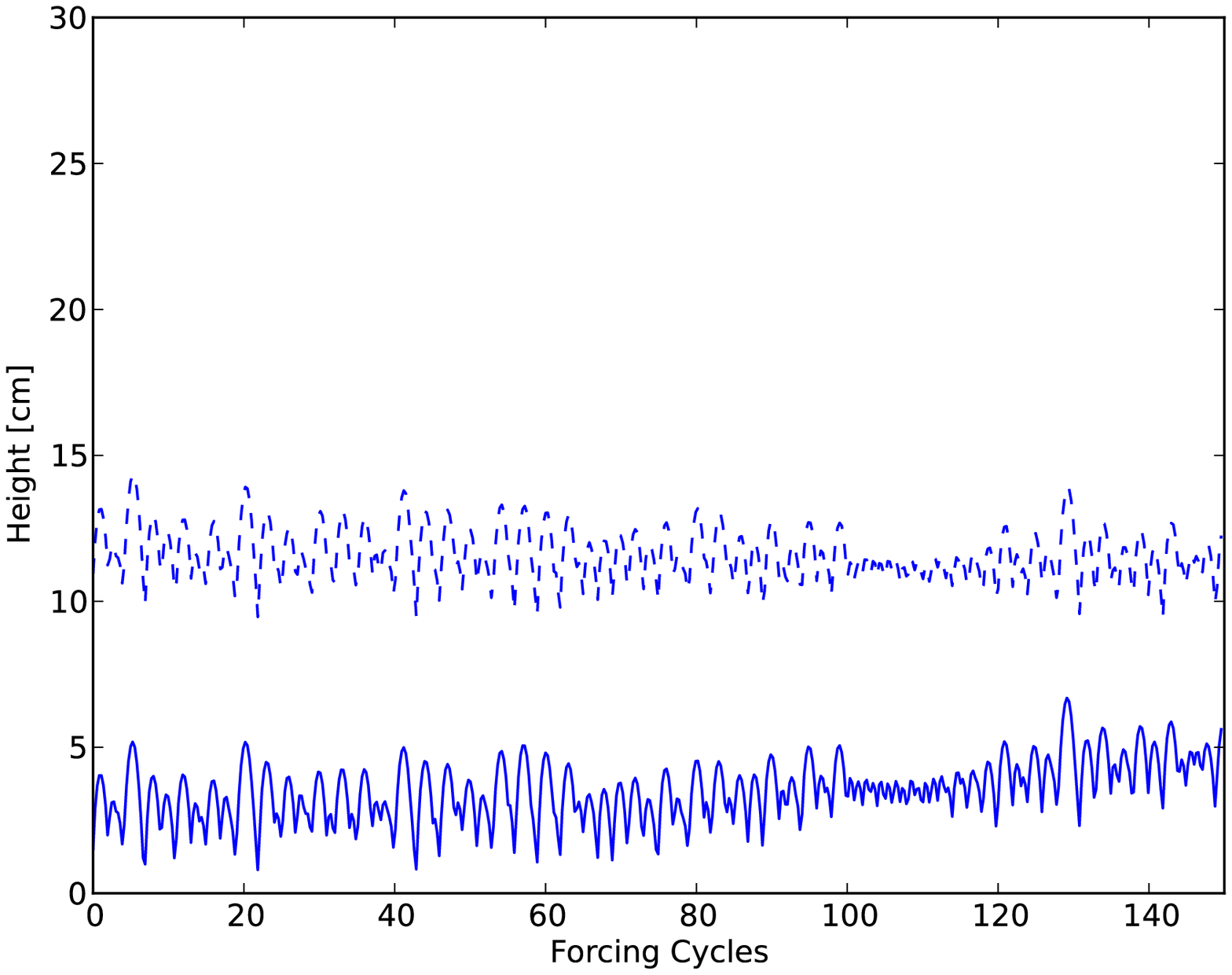}
\includegraphics[width=0.48\textwidth]{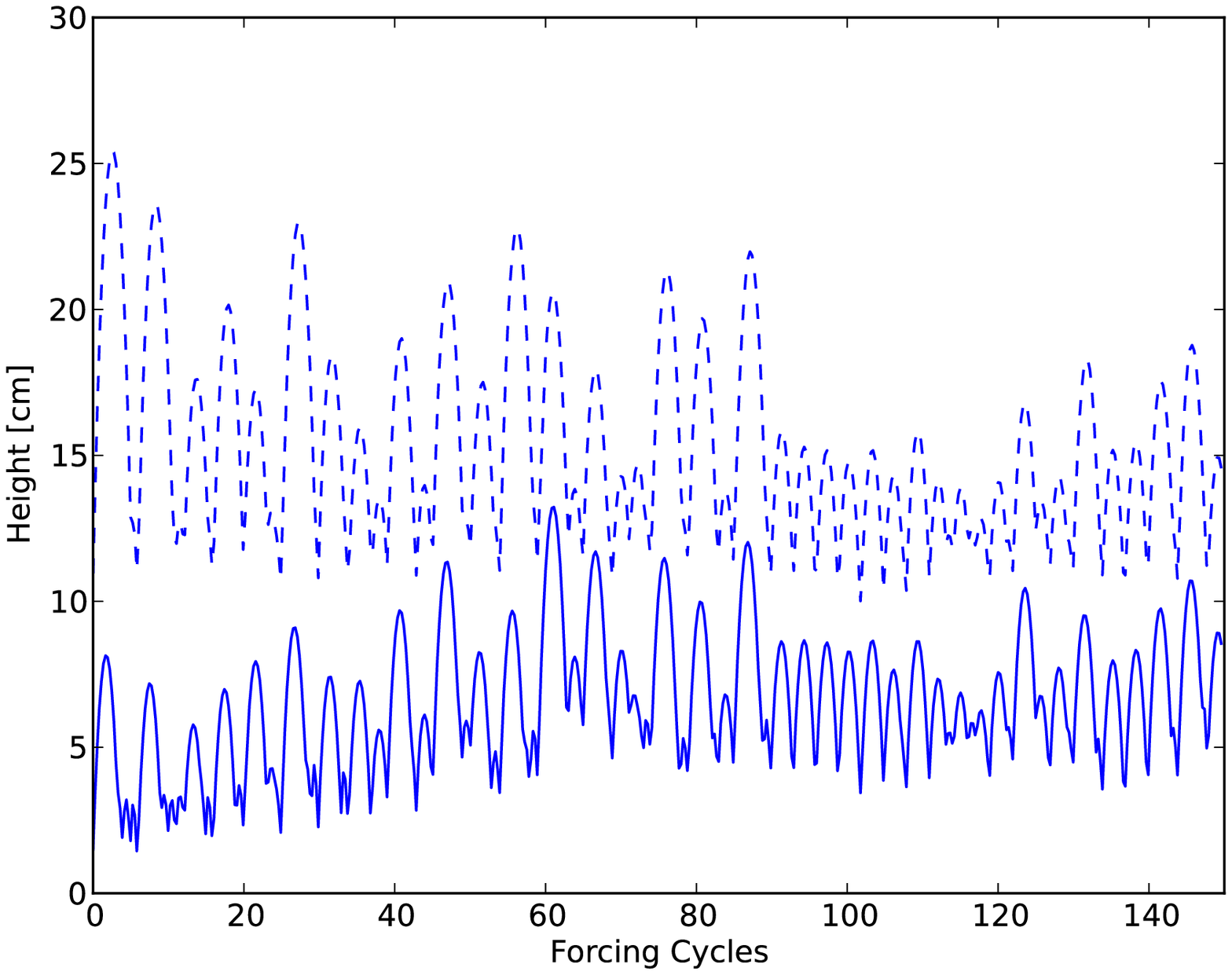}
\caption{Comparison of the
height evolution of the intruder with the median height of the small
particles. Left and right panels show the cases where particle-particle
frictions are set to zero (left) and particle-wall frictions are set to
zero (right), respectively. The other parameters are set to the usual default values. \label{fig_mu0}} 
\end{figure*}

Convection depends both on particle-particle and on particle-wall
frictions. Figure~\ref{fig_mu0} shows cases where particle-particle
(left) and particle-wall (right) friction constants are set to zero for
our default case. In both cases, convection does not occur and the BNE
is severely suppressed compared to the default case in 
Figure~\ref{fig_epsn0.5_epst0.5}, where both of these friction constants
have the default values. 
These figures indicate that kinetic friction 
(which is related to $\epsilon_n$ and $\epsilon_t$) alone is not enough to initiate the BNE.

Figure~\ref{fig_mus_mur} also indicates that high rolling friction diminishes the BNE. 
The higher rolling friction means that
it is more difficult for particles to rotate with respect to each other.
Let us consider three particles lined up side by side. If we move
the middle particle upward, then the two neighboring particles start
rotating. For higher rolling friction, the energy loss in this process
is greater and thus particles tend to lock to each other, which in turn 
slows down the convection. The slower rise of the intruder in the high-static-friction 
region is also due to the difficulty of the particles to move
relative to one another.

The friction constants of asteroids are even less constrained than the
coefficients of restitution. 
Recently, \shortcite{Yu2014inprep} performed avalanche experiments of similar-size gravels 
that were collected from a stream bed, and estimated the restitution coefficients and static 
constants by using numerical simulations on PKDGRAV.  
They found $\epsilon_n=\epsilon_t=0.55$, $\mu_s=1.31$ and $\mu_r=3.0$ reproduced 
their experiments well. 
These values are not necessarily unique for gravels, but the restitution coefficients are comparable to 
our default values while the static and rolling frictions are beyond the values we have 
investigated in this study (see Figure~\ref{fig_mus_mur}). If such gravels represent 
small particles in asteroids of interest, it would be very difficult to have convection and 
thus the BNE, because the particles will be simply locked to each other.  
However, their studies approximate gravels with spheres, and therefore could 
overestimate these values. More realistic modeling of particles would be necessary in 
the future. 
In the rest of the paper, we assume $\mu_s=0.7$ and
$\mu_r=0.1$ as our default values. Again, $\mu_s=0.7$ is comparable to
the friction constant estimated by \shortcite{Clement1992PhRvL} for oxidized
Al.
\shortcite{Yu2014inprep} also considered two other types of material: 
smooth ($\mu_s=\mu_r=0.0$) and glass ($\mu_s=0.43$ and $\mu_r=0.1$).  
Our default case has the higher static friction than glass but not as high as that of 
gravel.

\subsection{Dependence on Oscillation Speeds and Bed Depths}\label{results_vel} 
In this subsection, we study the dependence of the BNE on oscillation amplitude and frequency. 
For all the simulations in this subsection, we assume $\epsilon_n=\epsilon_t=0.5$,
$\mu_s=0.7$, and $\mu_r=0.1$.   
The oscillation speeds are varied for three different bed depths and two different 
cylinder widths.
The default case of 1800 + 1 particles in a cylinder with a diameter of 
$10\,$cm has a bed depth of $\sim22\,$cm. 
The shallower case of 900 + 1 particles has a depth of $\sim13\,$cm, 
and the deeper case of 3600 + 1 particles has a depth of $\sim47\,$cm in 
the same cylinder.  
We also performed one set of simulations in a wider cylinder with a diameter of 
$20\,$cm. That case is comprised of 3600 + 1 particles and a depth of $\sim13\,$cm; 
it can be compared with the shallow-bed case described above.

\begin{figure*}
\includegraphics[width=0.48\textwidth]{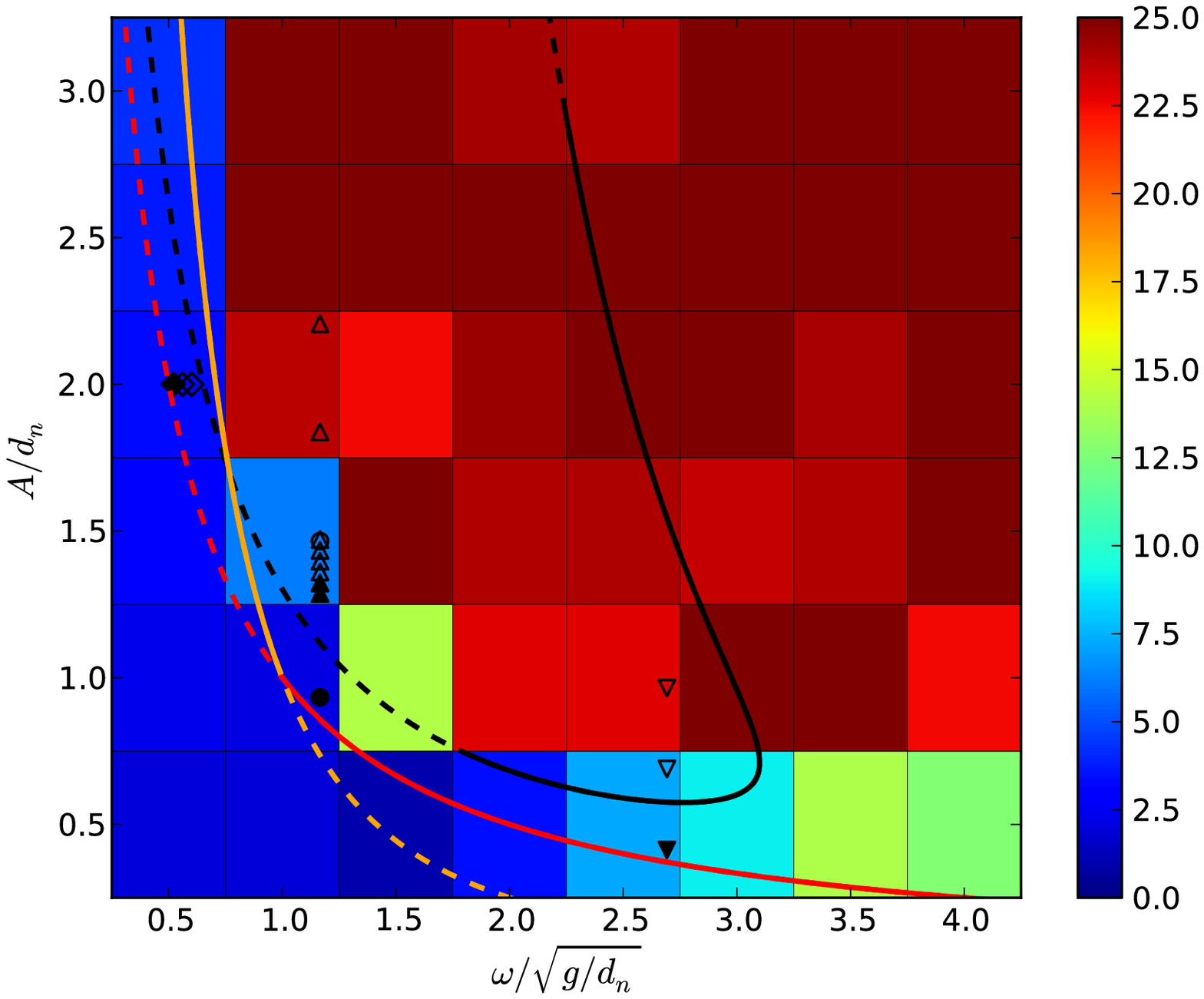}
\includegraphics[width=0.48\textwidth]{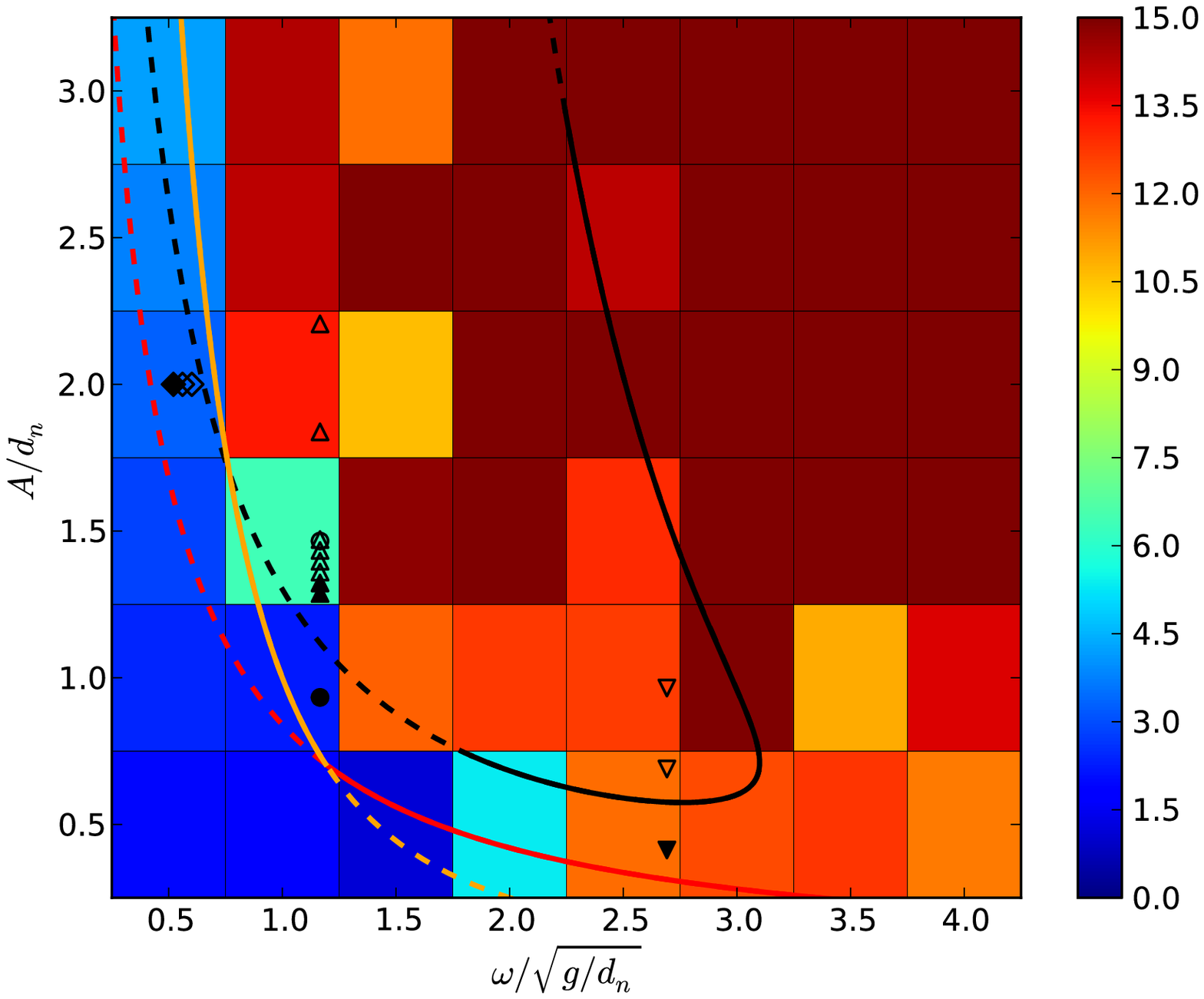}
\caption{
Final intruder heights after 150 oscillations as a 
function of $\tilde{\omega}$ and $\tilde{A}$.
Each rectangular region represents a specific choice of $\tilde{\omega}$ and $\tilde{A}$ and 
is color coded by the final intruder height (see color legend to the right of the plot).
For all the simulations, $\epsilon_n=\epsilon_t=0.5$, $\mu_s=0.7$, and $\mu_r=0.1$ are assumed. 
Left and right panels are the cases with $N_s=1800$ and $900$, respectively.
Interior to the black curve is the BNE region estimated from the 2D simulations by \shortcite{Godoy2008PhRvE}. 
The dashed part corresponds to the region beyond their investigation.
Open and filled symbols are taken from Figure~6 of \shortcite{Godoy2008PhRvE} and represent BNE and no-BNE cases, respectively. 
The circles correspond to the pseudo-2D experiments by \shortcite{Duran1994PhRvE}, while the downward triangles 
correspond to the 3D experiments by \shortcite{Knight1993PhRvL}.
The upward triangles and the diamonds are both 2D simulations by \shortcite{Saez2005PhRvE} and \shortcite{Poschel1995EL}, respectively. 
Orange and red curves are $\tilde{\Gamma}=1$ and $\tilde{v} = \tilde{v}_c$, respectively, 
and the convection is expected above the solid portions of these curves. 
 \label{fig_omega_A}}
\end{figure*}
The final heights of the intruders are plotted for
the dimensionless oscillation amplitude $\tilde{A}=0.5-3.0$ and 
the dimensionless oscillation frequency $\tilde{\omega}=0.5-4.0$ 
in Figure~\ref{fig_omega_A}, where  
\begin{equation} 
\tilde{A}\equiv \frac{A}{d_s},\quad                                                                          
\tilde{\omega}\equiv \frac{\omega}{\sqrt{g/d_s}} \ .
\end{equation}
Since the diameter of a small particle is $d_s=1\,$cm in our
simulations, our default case corresponds to $\tilde{A}=1$ and
$\tilde{\omega}=3$.
The left panel of Figure~\ref{fig_omega_A} shows our default case with 1800 + 1 particles and 
the right panel shows the shallower bed case with 900 + 1 particles.
The result for the deepest bed with 3600 + 1 particles (not shown here) looks similar to the default case.    
All of these simulations indicate that the BNE occurs for cases with sufficiently large amplitudes
and frequencies.
\begin{table*}
\footnotesize
 \begin{minipage}{180mm}
 \caption{Model Parameters and Setups \label{tab_init}}
 \begin{tabular}{lccccccc}\hline \hline 
 Reference & $N_s$ & $\rho_l/\rho_s$ & $d_l/d_s$ & Restitution coefficients & Friction constants & Type & Shake \\ \hline 
 Default case of present work & 1800 & 1 & 3 & $\epsilon_n=\epsilon_t=0.5$ & $\mu_s=0.7$, $\mu_r=0.1$ & 3DS & sinusoidal \\
 \shortcite{Hejmady2012PhRvE} & & $0.30-2.34$ & $8.5-11.3$ & & & 2DE & sinusoidal \\ 
 \shortcite{Godoy2008PhRvE} & 1200 & 1 & 8 & $\epsilon_n=\epsilon_t=0.98$ & $\mu_s=\mu_d=0.7$ & 2DS & parabolic \\ 
 \shortcite{Duran1994PhRvE} & & 1 & 12.9 & 0.5 & 0.8 & 2DE & sinusoidal \\
 \shortcite{Knight1993PhRvL} &  & 1 & $3-12.5$ &  &  & 3DE & sinusoidal taps \\
 \shortcite{Saez2005PhRvE} & 3300 &  & 13 & 0.6 & 0.97 & 2DS & sinusoidal \\ 
 \shortcite{Poschel1995EL} & 950 & 1 & $3.5-4.7$ &  & 0.5 & 2DS & sinusoidal \\ 
 \shortcite{Tancredi2012MNRAS} & 1000 & 1 & 3 & $\epsilon_n=0.8-0.9$, $\epsilon_t=0.95^*$ & $\mu_d=0.6$ & 3DS & displacements \\ 
 \hline \hline
 \end{tabular}
 \begin{tablenotes}
 \footnotesize
 \item{Column 2: number of small particles, Column 3: density ratio of large to small
 particles, Column 4: diameter ratio of large to small particles, Column
 7: dimension of simulations (S) or experiments (E), Column 8:
 oscillation type. The value with * is estimated from Figure 4 in \shortcite{Tancredi2012MNRAS}.}  
 \end{tablenotes}
 \end{minipage}
\end{table*} 
The figure also compares our results with the previous works listed in Table~\ref{tab_init}.
Since every work uses different setups, it is difficult to make a direct comparison.
One of the difficulties lies in a density dependence of the BNE.
In the vibro-fluidized regime, a high-enough density ratio $\rho_l/\rho_s$ could lead to the 
reverse BNE, where an intruder sinks to the bottom \cite[e.g.,][]{Ohtsuki1995JPSJ,Shishodia2001PhRvL}, 
while in the dense limit, when particles experience enduring contacts, 
an intruder appears to rise independent of the density ratio \cite[e.g.,][]{Liffman2001GM,Huerta2004PhRvL}.
Since we are interested in the standard BNE, we compare our models with previous works that assume comparable densities for 
small and large particles so that an intruder rises for both the vibro-fluidized regime and the dense limit.
     
\shortcite{Godoy2008PhRvE} selected several investigations that assume the same density for large and small particles, 
and showed that they all follow similar transition lines that separate BNE and no-BNE regions. 
These works are also plotted using different symbols in Figure~\ref{fig_omega_A}. 
The distribution of open (BNE) and filled (no BNE) symbols agrees well with the general trend of our simulations.

\shortcite{Duran1994PhRvE} experimentally studied the BNE in a quasi-two-dimensional bed of aluminum beads, 
and identified two segregation mechanisms depending on accelerations: arching ($1.0 \lesssim \tilde{\Gamma} \lesssim 1.5$) 
and convection ($\tilde{\Gamma} \gtrsim 1.5$), where $\tilde{\Gamma} = \tilde{A}\,\tilde{\omega}^2$ is the 
dimensionless acceleration. 
The orange line in Figure~\ref{fig_omega_A} corresponds to $\tilde{\Gamma} = 1$; 
the BNE is expected to take place to the right of this line.
The agreement is particularly good for a shallower bed case.
The default bed case, however, indicates that $\tilde{\Gamma} \gtrsim 1$ is not a sufficient condition for the BNE.  

\begin{figure}
\includegraphics[width=84mm]{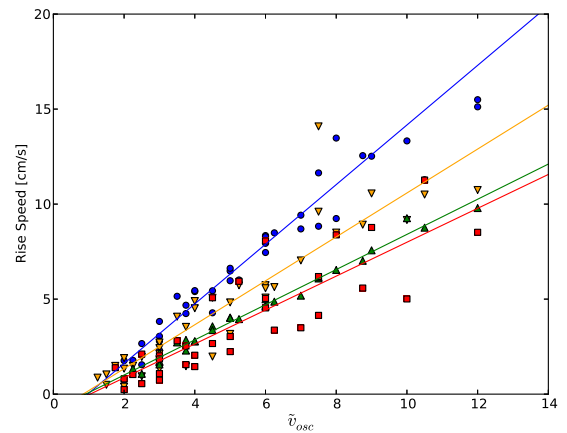}
\caption{Rise speed compared with the maximum oscillation speed. 
Blue circles, orange down-pointing triangles, and green up-pointing triangles correspond to the default case ($N=1800+1$), 
a shallow-bed case ($N=900+1$), and a deep-bed case ($N=3600+1$), respectively. 
Red squares represent a shallow-bed in a wider cylinder with $N=3600+1$.
All of these cases have a similar critical oscillation speed of $\tilde{v}_c\sim1$.
From the best-fit lines, the critical oscillation speeds are $\tilde{v}_c\sim0.97$, 
$0.84$, $0.91$, and $1.04$ for each case, respectively. See discussion in Section~\ref{results_vel}.
\label{fig_vosc_vrise}}
\end{figure}

\shortcite{Hejmady2012PhRvE} also experimentally studied the BNE by using a large
acrylic disk embedded in a quasi-two-dimensional bed of mustard seeds. 
They showed that $\tilde{\Gamma} > 1$ is not a sufficient condition for bulk convection to occur, 
and proposed that the oscillation speed also needs to exceed some critical value $v_{osc} > v_{c}$. 
We estimate the critical oscillation speed for our simulations in 
Figure~\ref{fig_vosc_vrise}, where the rise speed of an intruder is plotted 
as a function of the scaled maximum oscillation speed 
$\tilde{v}_{osc}=\tilde{A}\,\tilde{\omega}$ for three different bed depths and two 
different cylinder widths.
Here, the rise speed is defined as the bed depth (defined as $2\bar{z}_s$) 
divided by the rise time (determined from the rise cycle 
defined in Section~\ref{results_eps}).   
Different from \shortcite{Hejmady2012PhRvE}, we plot the rise speed instead of the rise time 
so that $\tilde{v}_c$ is clearly determined from the oscillation speed that 
corresponds to the zero rise speed.
We find that all of these cases have comparable critical speeds.
For our default bed depth, we find $\tilde{v}_c\sim0.97$, while 
for shallower and deeper bed cases, we find $\tilde{v}_c\sim0.84$ and $0.91$, respectively.  
For the case of a shallow-bed in a wide cylinder, we find $\tilde{v}_c\sim1.04$.
These are comparable to the critical value found in \shortcite{Hejmady2012PhRvE} 
($\tilde{v}_c\sim1.26$ for $v_c=16.5\,{\rm cm \, s^{-1}}$).
The estimated critical speeds are plotted using red lines in Figure~\ref{fig_omega_A}.
From these figures, we conclude that both $\tilde{\Gamma}\gtrsim1$ and $\tilde{v}_{osc}\gtrsim\tilde{v}_c$ 
need to be satisfied for the BNE to take place.  

An important indication from Figure~\ref{fig_vosc_vrise} is that 
the rise speed is proportional to the oscillation speed. 
We will come back to this point in Section~\ref{results_lowg}.
Furthermore, Figure~\ref{fig_vosc_vrise} indicates that there is an optimal bed depth 
for the BNE, since the rise speed 
increases from $N_s=900$ to $1800$, and then decreases from $N_s=1800$ to $3600$.
Such a depth-dependence of the rise time agrees with the recent experimental result by \shortcite{Guttler2013PhRvE}.   

Comparison of the shallow bed case in the default cylinder to that in the wide cylinder 
(i.e., orange and red lines, respectively) indicates that the BNE may take place slower in a wider cylinder. 
For the wide-cylinder cases, we find that the convection direction is the same as the default cylinder cases 
--- the particles descend along the wall and ascend near the central region.
For the high-end of oscillation speeds, however, we find that the intruder also shows a ``whale'' effect, 
where the intruder does not necessarily stay at the top of the cylinder but keeps moving up and down with the convective current. 
This indicates that the region of the convection roll along the wall is thick enough to pass the intruder 
downward in the wide-cylinder cases.

We would like to pay a particular attention to the comparison of our work with \shortcite{Godoy2008PhRvE}. 
They numerically studied the BNE by using the molecular
dynamics code developed by \shortcite{Risso2005PhRvE}. In their simulations,
the size ratio of the intruder and surrounding disks is $d_l/d_s=8$,
restitution coefficients $\epsilon_n=\epsilon_t=0.98$, and static and
dynamic friction coefficients $\mu_s=\mu_d=0.7$. 
Here, $\mu_d$ follows the Coulomb's law of friction where the magnitude of kinetic friction 
is independent of speed of slippage.
There are 1200 small
particles with one intruder, and the 2D box has a width of $40\,d_s$.
Their oscillations are not sinusoidal, but are given in a periodic
vertical parabolic movement with the base acceleration of $\pm
\frac{8}{\pi^2}A\,\omega^2$. 
Since their acceleration is proportional to the maximum acceleration 
of the sinusoidal oscillation $A\,\omega^2$, 
we expect that our results can be compared to theirs reasonably well.

Instead of changing oscillation amplitude
and frequency like we have done in our work, \shortcite{Godoy2008PhRvE} varied the dimensionless acceleration and speed which
they defined as $\Gamma'=\frac{8}{\pi^2}\,\tilde{A}\,\tilde{\omega}^2$ and
$\zeta'=\sqrt{2}\,\tilde{A}\,\tilde{\omega}$, respectively, and found the transition
line above which the BNE is observed. We also plot their transition line
as the black line in Figure~\ref{fig_omega_A}. 
The agreement between their simulations and ours is good in the low-frequency region 
($\tilde{\omega}\lesssim2.5$), but our results disagree in the higher frequency region.

There are several possibilities that lead to the difference between our results and 
those of \shortcite{Godoy2008PhRvE}.
First, our simulations are three-dimensional (3D), while theirs are two-dimensional (2D).
However, this is unlikely to be the critical difference, especially since our 3D results agree well with 
quasi-2D experimental results by \shortcite{Hejmady2012PhRvE}.  
Second, we adopt $\epsilon_n=\epsilon_t=0.5$, while they use $\epsilon_n=\epsilon_t=0.98$.
\shortcite{Kudrolli2004RPPh} proposed a condition that separates the vibro-fluidized regime 
from the dense regime: $n_{\rm layer}(1-\epsilon) < 1$, where $n_{\rm layer}$ is the number 
of particle layers.
According to this condition, our simulations have $n_{\rm layer} \sim 22$ and $\epsilon=0.5$ and 
thus are likely to be in the dense regime of lasting contacts between particles, while 
their simulations have $n_{\rm layer} \sim 30$ and $\epsilon=0.98$ and 
may be in the vibro-fluidized regime, especially for high accelerations.   
To address this, we repeated our simulations for $\epsilon_n=\epsilon_t=0.98$. 
However, the results are still consistent with $\tilde{\Gamma}\gtrsim1$ and $\tilde{v}_{osc}\gtrsim\tilde{v}_c$, 
rather than with the relation proposed by \shortcite{Godoy2008PhRvE}.
Third, it is possible that our oscillation speeds are too low to turn the BNE off. 
Thus, we extended our default simulations up to $\tilde{\omega}=8$ and $\tilde{A}=4$.      
However, the BNE is observed for all oscillation speeds we tested.   
The disagreement may also be due to the differences between our codes, or other differences between our initial setups.    
Future studies would need to investigate, both numerically and experimentally, 
whether the BNE turns off for high oscillation frequencies or not. 

In summary, we have found that the BNE occurs for the oscillation frequency and oscillation 
amplitude above certain values, and that critical conditions are well approximated by 
$\tilde{\Gamma}\gtrsim1$ and $\tilde{v}_{osc}\gtrsim1$ for the parameters we tested. 
Our results show the same general trend as other works that use comparable densities for small and large particles 
but assume different initial conditions otherwise.
In Section~\ref{results_lowg}, we investigate the effects of oscillations under the low-gravity environments, 
and discuss a possibility of having such oscillations due to impact-generated seismic waves.

\subsection{Comparison with Tancredi et al. (2012)}\label{results_Tancredi12} 
We compared our simulations with previous works listed in Table~\ref{tab_init} in the last subsection.
We left out the results of \shortcite{Tancredi2012MNRAS} from this discussion due to the very
different oscillation style they used. In this subsection, we use initial
conditions as close to theirs as we could construct, and then compare the results.

\shortcite{Tancredi2012MNRAS} studied granular processes under various
gravitational environments, and observed size segregation in response to shaking. 
Instead of the sinusoidal oscillation, they applied 
multiple vertical displacements of the floor of the simulation box at a
constant speed. The duration of a displacement is $dt=0.1\,$s, and the
time between displacements varies from $2-15\,$s, depending on the
gravitational environments. The floor's speed ($v_f=0.3-10\,{\rm m \, s^{-1}}$) is
linearly increased from 0 to the final value in 20 displacements. 
To mimic their oscillations, we do not use the vertical displacements, but instead
increase the oscillation amplitudes linearly so that the maximum 
speeds reach their final floor speeds in 20 cycles. The oscillation
amplitude is chosen to be equal to one displacement $A=v_f\,dt$. This
makes the oscillation frequency $\omega=10\,{\rm rad \, s^{-1}}$ for all the
simulations.

Their simulation box has a size of $12\,$m$\times12\,$m and a height of $150\,$m. Due
to the lack of viscoelastic and frictional interactions between
particles and walls in their ESyS-PARTICLE code, they glued a layer of
small particles on the floor. Since our code handles the wall-particle
interactions, we do not have the glued layer.

Following their setup, we create an infinitely-tall box with the floor area of 
$12\,$m$\times12\,$m and fill that box with 1000 small spheres with mean radius $0.25\,$m
and standard deviation $0.01\,$m as well as one large sphere with radius $0.75\,$m. 
The restitution coefficients are estimated from
their sections 2.3.1--2.3.3 as follows. In the head-on collision simulation of two
equal spheres (see their section~2.3.1), they obtained restitution
coefficients of $0.8-0.9$ for the impact speeds over $1-2\,{\rm m \, s^{-1}}$.
Thus, we adopt $\epsilon_n=0.85$ for the normal coefficient of
restitution between particles. In the grazing collision simulation of
two equal spheres (see their section~2.3.2), they found that the ratio of 
final to initial speeds is $\sim0.95$ for the low-speed collisions with friction.
Thus, we adopt $\epsilon_t=0.95$ for the tangential coefficient of
restitution between particles. In the bouncing ball simulation against
the floor with a layer of glued spheres (see their section~2.3.3), they found a coefficient of
restitution of 0.593. We adopt $\epsilon_n=\epsilon_t=0.593$ not only for 
the interactions between particles and the floor, but those for all the
walls of the box.

The treatment of friction is also different between the codes. We
chose $\mu_s=0.6$ so that our friction constant becomes comparable
to theirs at the threshold between static and dynamic friction. Since their code
does not take account of rolling friction, we set $\mu_r=0.0$.

\begin{figure}
\includegraphics[width=84mm]{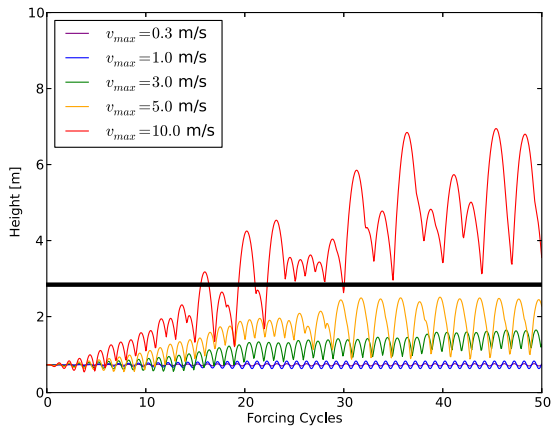}
\caption{Evolution of the height of the intruder for the maximum
oscillation speeds of 0.3, 1, 3, 5, and $10\,{\rm m \, s^{-1}}$. A thick black line is
drawn at a height of $2.84\,$m to represent the height of 1000 small
particles with a random close packing and maximum bulk porosity of 0.64.
The figure qualitatively agrees with Figure 9 in Tancredi et al. (2012).
 \label{fig_Tancredi}} 
\end{figure}

With this setup, we performed oscillation simulations for the maximum
speeds of 0.3, 1, 3, 5, and $10\,$m/s and show the results in
Figure~\ref{fig_Tancredi}. The figure is meant to be compared with
Figure~9 in \shortcite{Tancredi2012MNRAS}. Despite the differences between
our codes and shaking styles, our results qualitatively agree with
theirs, apart from a difference in rise time. 
For the low oscillation speeds (0.3 and $1\,{\rm m \, s^{-1}}$), the intruder
stays at the bottom of the box. As the maximum oscillation speed
increases, the intruder starts rising, showing the familiar BNE. 
The duration of the maximum speed in our oscillation simulation (a
fraction of the oscillation period of $0.63\,$s) is shorter than theirs.
Since the rise speed is proportional to the oscillation speed (see Section~\ref{results_vel}), 
it is understandable that our results show a consistently slow rising 
time for the intruders compared to \shortcite{Tancredi2012MNRAS}. 

\subsection{Scaling in Low-gravity Environments}\label{results_lowg}
In this subsection, we apply our model to the low-gravity environments 
characteristic of asteroids. 
The goal here is to check whether the BNE occurs in such environments, and 
to understand the gravity dependence of the rise time of an intruder.

We expect that the BNE is scalable by adjusting the oscillation frequency according 
to the gravity field. 
For example, our default case has an oscillation frequency of
$\omega=3\sqrt{a_g/d_s}=93.9\,$rad/s under the
gravitational acceleration of Earth $a_g=g=980\,{\rm cm \, s^{-2}}$.
We change the oscillation frequencies for the gravity fields comparable to the Moon,
(1) Ceres, (87) Sylvia, Eros, and Itokawa accordingly. 
These bodies are chosen to compare our results with those in \shortcite{Tancredi2012MNRAS}.

We tested four different parameter sets --- $(\tilde{\omega},\, \tilde{A}) = (1,\, 0.5)$, $(1,\, 1)$, $(3,\, 1)$, and $(3,\, 3)$.
As expected, we find that the results look similar for different gravity environments with the same scaled oscillation speeds.
For $(\tilde{\omega},\, \tilde{A}) = (1,\, 0.5)$ and $(1,\, 1)$, the BNE does not occur for any gravity fields, 
while for $(\tilde{\omega},\, \tilde{A}) = (3,\, 1)$ and $(3,\, 3)$, the BNE occurs for all the cases.

\begin{figure}
\includegraphics[width=84mm]{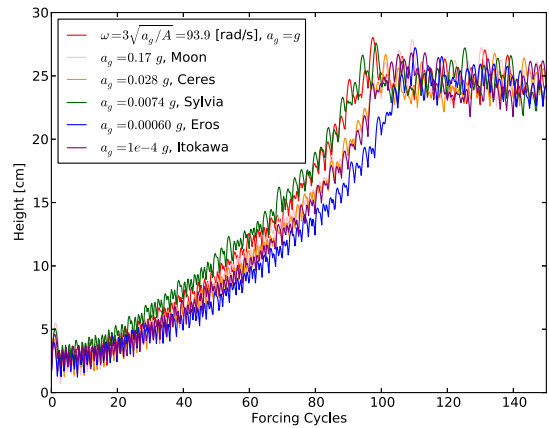}
\caption{Evolution of height of the intruder as a function of
oscillation cycles for the gravities of Earth, Moon, Ceres, Sylvia,
Eros, and Itokawa.  
\label{fig_lowg}} 
\end{figure}

The comparison of the evolution of an intruder's height for $(\tilde{\omega},\, \tilde{A}) = (3,\, 1)$ 
is plotted in Figure~\ref{fig_lowg}.  
We find that the results are all consistent with
the uncertainties typical to the oscillation simulations. 
The figure confirms the expectation that the BNE simulations are scalable over a wide range
of gravity fields.

\begin{figure} 
\includegraphics[width=84mm]{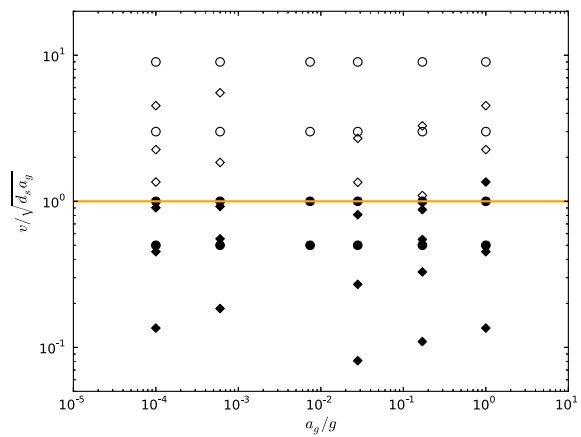}
\caption{The scaled oscillation speeds compared with the scaled gravitational acceleration. 
The solid orange line corresponds to the critical speed $\tilde{v}_c=1.0$ estimated in Section~\ref{results_vel}.
Open and filled symbols correspond to BNE and no-BNE cases in our simulations (circles) and in \shortcite{Tancredi2012MNRAS} (diamonds). 
For data from \shortcite{Tancredi2012MNRAS}, we simply follow their Figure~11 to distinguish BNE and no-BNE cases.
\label{fig_grav_vel}} 
\end{figure}

In Figure~\ref{fig_grav_vel}, we compare our simulations under different gravity fields and the maximum oscillation speeds.
The open and filled circles correspond to the runs with and without the BNE, respectively.
The orange line is the critical speed we estimated in Section~\ref{results_vel}. 
As expected, BNEs are observed above $\tilde{v}_c$, while BNEs do not occur below the critical value.

For a comparison, we also plot BNE (open diamonds) and no-BNE cases (filled diamonds) 
from low-gravity simulations in \shortcite{Tancredi2012MNRAS}. 
Since their results also show a similar transition from no BNE to BNE around a constant $\tilde{v}_c$, 
we would expect that there is a critical floor speed necessary for the BNE in their simulations as well. 
\shortcite{Tancredi2012MNRAS} estimated the floor speed threshold as $v_{\rm thre}=1.12 a_g^{0.42}$, 
which has a slightly different dependency on the gravitational acceleration from our relation of $v_{c}\sim \sqrt{d_s a_g}$. 
This difference may not be surprising since threshold speeds in \shortcite{Tancredi2012MNRAS} 
are determined by relatively sparse data.

\begin{figure*}
\includegraphics[width=0.48\textwidth]{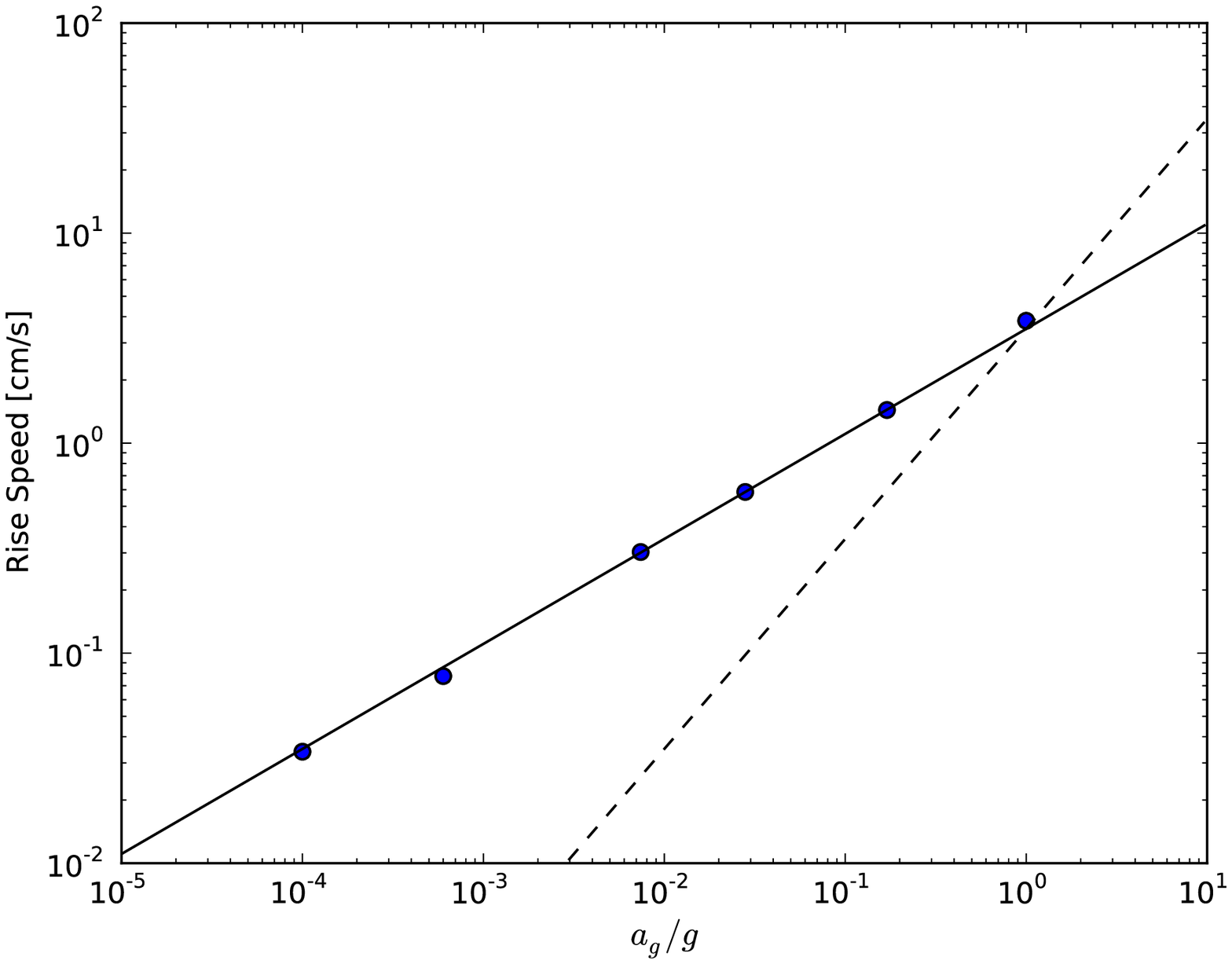}
\includegraphics[width=0.48\textwidth]{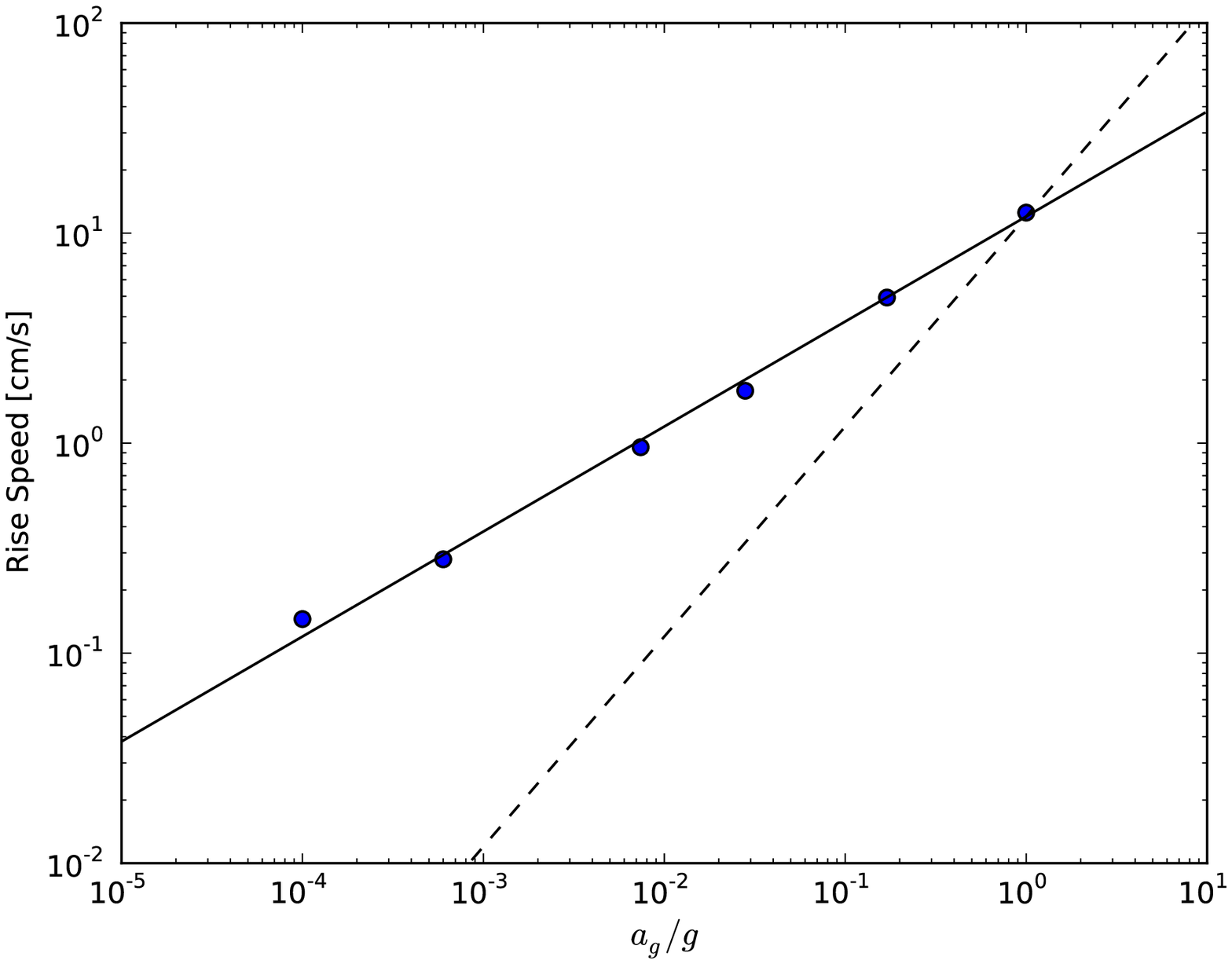}
\caption{Change of the rise speed of an intruder depending on the gravity field. 
Left and right panels correspond to $(\tilde{\omega},\, \tilde{A}) = (3,\, 1)$, and $(3,\, 3)$, respectively. 
Solid and dashed lines are proportional to $\sqrt{a_g}$ and $a_g$, respectively.
The rise speeds of both sets of our simulations have a $\sqrt{a_g}$ dependence.  \label{fig_grav_vrise}}
\end{figure*}

It is also informative to plot the rise speed of an intruder as a function of the gravity field.
Figure~\ref{fig_grav_vrise} shows that, for both $(\tilde{\omega},\, \tilde{A}) = (3,\, 1)$ and $(3,\, 3)$, 
the rise speed is proportional to $\sqrt{a_g}$, rather than $a_g$.
This is understandable in our case, since the driving frequency of our oscillation simulations is $\omega \propto \sqrt{a_g}$.  
The result is consistent with our finding in Section~\ref{results_vel} 
that the rise speed is proportional to the maximum oscillation speed. 
  
Recently, \shortcite{Guttler2013PhRvE} experimentally studied the BNE both in the laboratory and in a parabolic flight to mimic the 
reduced-gravity conditions comparable to Mars and the Moon.  
They found that the rise speed was not proportional to $\sqrt{a_g}$, but closer to $a_g$.
In fact, their best fit was obtained for an exponential function.
The difference seen in our rise speed dependences with gravity may be partly due to the difference in our shaking profiles.
While we use a sinusoidal oscillation, their oscillation acceleration is approximated by a square-wave function.  
Moreover, due to the nature of a parabolic flight, they need to stop the oscillations every time the hyper-gravity phase kicks in, 
which could have compacted the particle distributions and slowed the rise of an intruder. 
Future studies need to investigate the dependence of rise speed (or equivalently rise time) on the gravity field further.

\begin{figure*}
\includegraphics[width=0.48\textwidth]{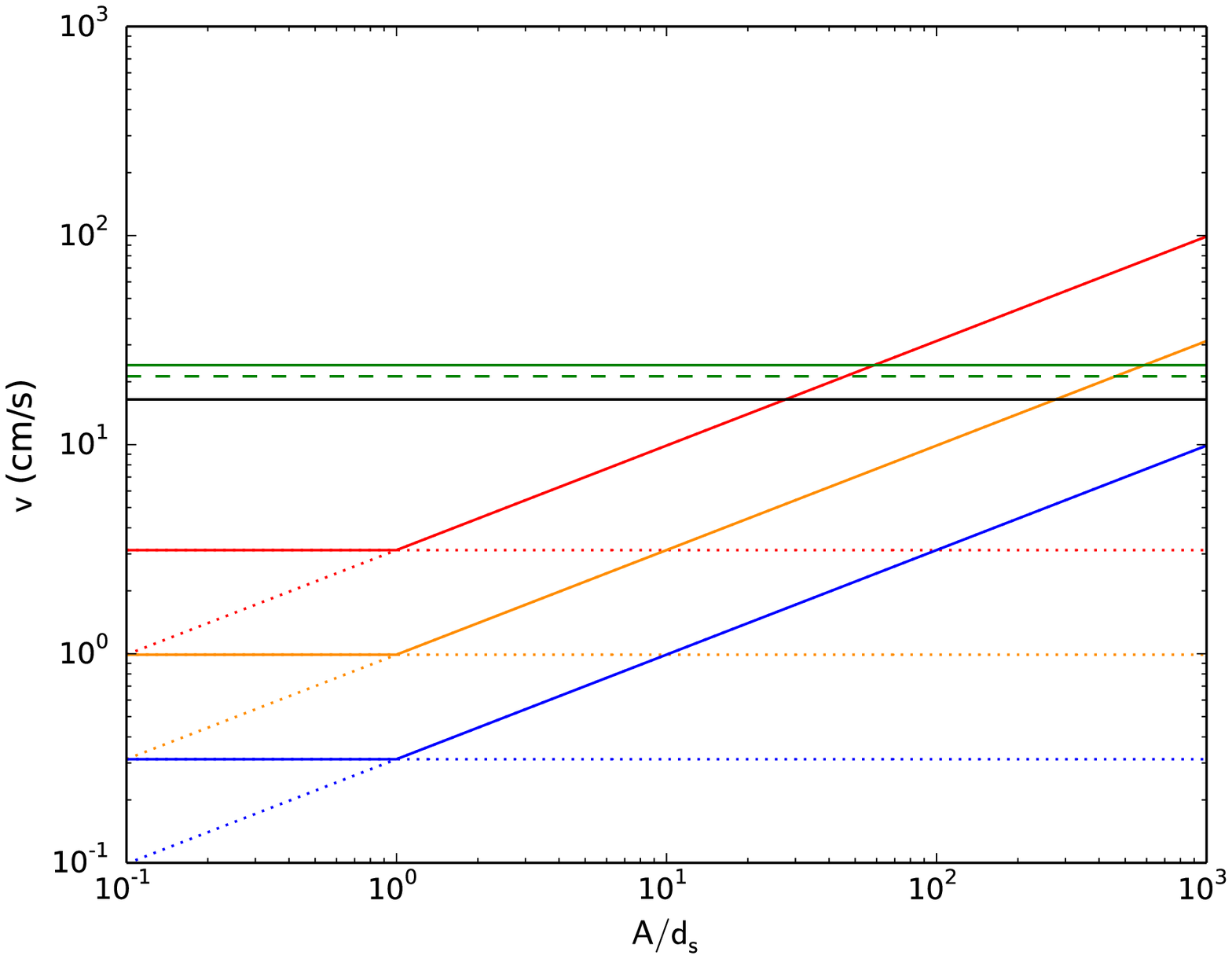}
\includegraphics[width=0.48\textwidth]{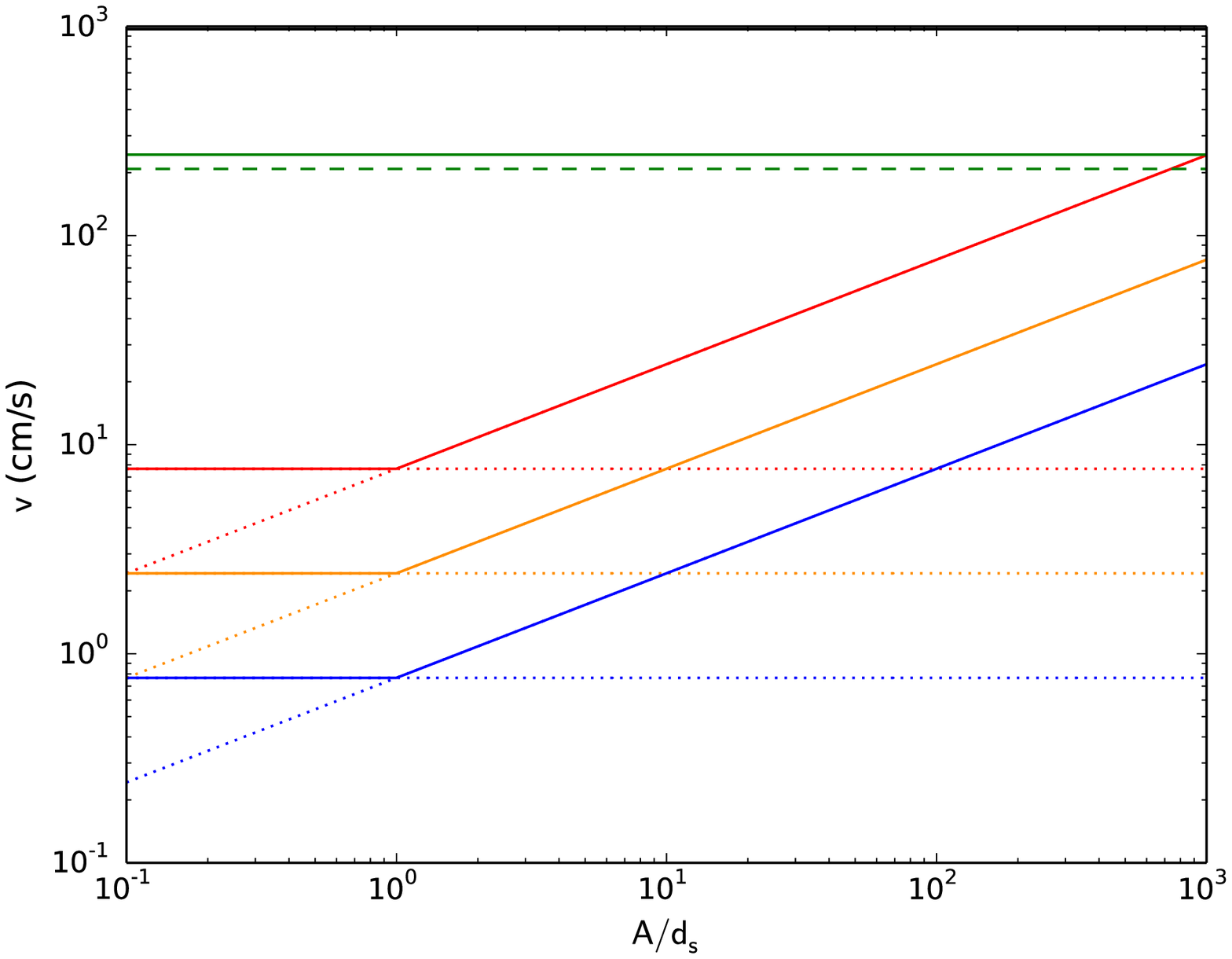}
\caption{The critical oscillation speeds necessary for the BNE that is estimated from 
$\tilde{\Gamma}_c=1.0$ and $\tilde{v}_c=1.0$ for Itokawa (left panel) and Eros (right panel). 
Blue, orange, and red lines correspond to a typical small particle's size of 
1, 10, and 100\,cm, respectively. 
Green solid and dashed lines are the seismic speeds estimated from an impact that 
lead to a catastrophic disruption, 
and correspond to the impact speeds of $5\, {\rm km \, s^{-1}}$ and $3\, {\rm km \, s^{-1}}$, respectively.    
Larger projectiles would destroy asteroids (above these lines), while smaller ones would generate 
seismic speeds (below these lines) that are comparable to the critical oscillation speeds for the BNE. 
The solid black line is the escape speed.
 \label{fig_vcrit}}
\end{figure*}

Finally, we estimate the critical oscillation speeds of the BNE for observed asteroids. 
From the critical conditions confirmed in Section~\ref{results_vel}, 
we can derive two conditions for critical oscillation speeds, 
\begin{eqnarray}
\tilde{v}_c\gtrsim 1.0 & \rightarrow & v \gtrsim \sqrt{d_s a_g} \ {\rm, \, and} \\
\tilde{\Gamma}_c\gtrsim 1.0 & \rightarrow & v \gtrsim \sqrt{A a_g} \ .
\end{eqnarray}
In Figure~\ref{fig_vcrit}, we plot these conditions for Itokawa (left panel) and Eros (right panel) 
by assuming three different diameters for a typical small particle (1, 10, and 100\,cm).
When the particle size is $d_s=10\,$cm, the oscillation speeds need to be larger than $v_c\sim1\,{\rm cm \, s^{-1}}$ 
for the BNE to take place on Itokawa, and larger than $v_c\sim2.5\,{\rm cm \, s^{-1}}$ on Eros.
Thus, as expected from the relation $v_c \propto \sqrt{a_g}$, size segregations occur for weaker oscillations 
on smaller asteroids.

The upper limit of the BNE oscillation speeds can be set by the escape speed. 
The escape speeds are $\sim16.5\,{\rm cm \, s^{-1}}$ for Itokawa and $\sim9.71\,{\rm m \, s^{-1}}$ 
for Eros, and are plotted as black solid lines in Figure~\ref{fig_vcrit} 
(note that the latter line overlaps with the top boarder of the panel).
For a small asteroid like Itokawa that has a diameter $< 500\,$m, 
the BNE is allowed only for small oscillation amplitudes $\lesssim 10\,$m if a typical small particle size is 100\,cm.
For smaller particles, a wider range of conditions is possible.
For a larger asteroid like Eros that has a diameter of $\sim20\,$km, 
a typical small particle size can be larger than 100\,cm. 

How do these oscillation speeds compare to the characteristic seismic speeds in asteroids?
From \shortcite{Asphaug1996Icar}, the estimated critical oscillation speeds are consistent 
with the speeds of seismic motions that could create bright annuli around craters on small asteroids such as Ida.
We estimate the critical seismic speeds below and plot them in Figure~\ref{fig_vcrit} as 
solid and dashed green lines.
 
The impact energy of a projectile can be written as
\begin{equation}
E_i = \frac{2\pi}{3}\rho_p R_p^3 v_p^2 \ ,
\end{equation}
where $p$ denotes the projectile, 
$\rho$ is the density, $R$ is the spherical radius, and $v_p$ is the impact speed.
Following \shortcite{Richardson2005Icar}, we also write the total seismic 
energy of an asteroid due to a seismic wave speed $v_s$ as
\begin{equation}
E_s = \frac{2\pi}{3}\rho_t R_t^3 v_s^2 \ .
\end{equation}
Here, $t$ denotes the target asteroid and $v_s=2\pi f A$, 
where $f$ and $A$ are the seismic frequency and maximum displacement, 
respectively, from \shortcite{Richardson2005Icar}.
More complex representations may be 
appropriate for purely agglomerate bodies, but since there is no good
understanding of the actual structure of agglomerate asteroids or 
the transmission of seismic waves within such a medium, we have assumed that
the waves are transmitted in a simplified fashion (i.e., a sinusoidal oscillation).

When a fraction $\eta$ of the kinetic energy of an impactor 
is converted into seismic energy (i.e., $E_s=\eta E_i$),
the seismic speed $v_s$ on an asteroid is estimated as
\begin{equation}
v_s = \sqrt{\eta \frac{\rho_p}{\rho_t}\left(\frac{R_p}{R_t}\right)^3} v_p
\end{equation}
%
By defining a specific impact energy as  
\begin{equation}
Q_{s} = \frac{1}{2}\frac{\rho_p}{\rho_t}\left(\frac{R_p}{R_t}\right)^3 v_p^2 \ , 
\end{equation}
we can rewrite a seismic speed as $v_s = \sqrt{2\eta Q_{s}}$.

To estimate a seismic speed for a certain impact, we consider an extreme case 
where an impact leads to a catastrophic disruption (i.e., an impact that results in the largest 
remnant having half the mass of the original target) 
and define the specific impact energy as $Q_s=Q_{s,D}$.
\shortcite{Jutzi2010Icar} performed numerical simulations of asteroid break-ups, 
and estimated such specific energy threshold for disruption $Q_{s,D}$.
They provided the following convenient power-law scaling:  
\begin{equation}
Q_{s,D} = Q_0 \left(\frac{R_t}{\rm cm}\right)^\alpha 
            + B\rho_t\left(\frac{R_t}{\rm cm}\right)^\beta  \ ,
\end{equation}
where $Q_0$, $B$, $\alpha$, and $\beta$ are fitting constants.
The first term represents the strength regime where fragments are held together by material 
strength alone, and the second term represents the gravity regime 
where fragments are gravitationally bound.
The transition between these two regimes is estimated to occur for target diameters 
between 100 and 200\,m \cite[][]{Benz1999Icar}.
The fitting constants depend on a variety of parameters such as internal structures, 
tensile and shear strengths of constituents, and impact speeds. 
\shortcite{Jutzi2010Icar} considered two models for the target's internal structure --- 
a purely monolithic non-porous target and a porous target that consists of a body with pores 
that are smaller than the thickness of the shock front, and determined 
the fitting constants for impact speeds of $3$ and $5\, {\rm km \, s^{-1}}$.
Since both Itokawa and Eros have low bulk densities compared to ordinary chondrites 
\cite[][]{Wilkison2002Icar,Fujiwara2006Sci}, we adopt the fitting parameters for a porous target.
By assuming impact seismic efficiency of $\eta = 10^{-4}$ 
\cite[][and references therein]{Richardson2005Icar}, 
we plot critical seismic speeds $v_s$ for impact speeds of $3$ and $5\, {\rm km \, s^{-1}}$ 
as dashed and solid green lines, respectively, in Figure~\ref{fig_vcrit}.   
Regions below these lines correspond to seismic speeds expected for smaller 
impactors that will not destroy the asteroids.

For Itokawa, we find that critical seismic speeds would be comparable to, 
but slightly larger than, the escape speed, and 
larger than the minimum oscillation speeds required for the BNE.
For Eros, critical seismic speeds would be smaller than the escape speed, but 
they are still larger than the estimated minimum oscillation speeds.  
Thus, in both cases, we expect that the critical BNE oscillation speeds are comparable to the 
seismic speeds that can create craters.

It is also informative to speculate how long it might take for a large block to rise to the surface of an asteroid.
From Figure~\ref{fig_vosc_vrise}, we find that the rise speed of an intruder is 
more than an order of magnitude slower than the maximum oscillation speed for these three bed depths.
Assuming an oscillation speed of $1\,{\rm cm \, s^{-1}}$ and a depth of 100\,m from the shortest axis of Itokawa of $\sim200\,$m, 
we can estimate that the rise time would be a few hours if the rise speed is comparable to the oscillation speed, 
and more than a day if the rise speed is one tenth of the oscillation speed.
Similarly, by assuming an oscillation speed of $100\,{\rm cm \, s^{-1}}$ and a depth of 5.5\,km from the shortest axis of Eros of $\sim11\,$km, 
the rise time would be about 90\,minutes for the oscillation speed, and about 15\,hours for one tenth of that speed. 
This implies that unless the seismic shaking lasts for more than a couple of hours, 
one impact might not be enough for a large block to rise to the surface, 
and that multiple impacts would be necessary to change the surface significantly.

A potential problem for size sorting by multiple impacts is that 
the oscillation directions will likely vary for different impacts.
To check this point, we performed simulations where we changed the direction and magnitude 
of {\it effective} gravity instead of oscillating cylinders sinusoidally.
First, we consider the most ideal case where the impact is always applied in the vertical direction of the cylinder.
We use the initial conditions identical to the default case, and change only the direction of the gravity as follows.
Initially, the gravity is in the $-z$ direction with the magnitude of the Earth's gravity.  
Once the simulation starts, we apply a ``jolt'' to the cylinder by changing the gravity vector first to the $+z$ direction 
and then back to the $-z$ direction within a short period of time that is randomly chosen from $0.1-0.3\,$s.
We then keep on applying the Earth gravity in the $-z$ direction for a randomly chosen period of time from $0.3-0.4\,$s.
After this break, we again apply a jolt as described above, and repeat these steps a number of times.
Here, the choice of the duration of the jolts is arbitrary.
In this set up, we find that the intruder comes to the surface within $\sim20\,$s, 
which is comparable to the rise time of the low-speed oscillation (e.g., $\tilde{\omega}=1.0$ and $\tilde{A}=2.0$).
Next, we consider a less ideal case where the impact is applied in a random direction relative to the cylinder.
Similar to the above case, the cylinder is initially under Earth's gravity in the $-z$ direction.
To mimic a jolt, we randomly choose the gravity magnitude from $-2g$ to $2g$ in each direction of $x$, $y$, and $z$, and smoothly change 
the magnitude and direction of the gravity vector from and then back to the initial one within $0.1-0.3\,$s.
After the break of $0.3-0.4\,$s, we apply another random jolt and repeat these steps a number of times.
We find that the rise of the intruder is much slower in this setup.   
After $\sim180\,$s, the intruder is about one quarter of the bed depth from the bottom of the cylinder.  
Therefore, we expect that the rise time would be much longer than the ones estimated from our sinusoidal oscillations 
if the oscillations are applied from random directions.
If, on the other hand, the oscillations are applied approximately perpendicular to a particle bed, 
the rise time would be less affected.
%
%
%
\section{Discussion and Conclusions}\label{summary} 
We have studied the BNE in an intruder model by using the $N$-body code PKDGRAV
with a soft-sphere collision model, and explored its effect on asteroids.
We have also investigated the dependence of the BNE on the particle properties, 
oscillation conditions, and gravitational environments. 

Similar to previous studies, we have found that convection is the major driving 
source for the BNE. 
Our main conclusions are the following.
\begin{enumerate}
\item{The occurrence of the BNE as well as the rise time of an intruder is largely independent of the choice 
of coefficients of restitution (see Section~\ref{results_eps}). For highly elastic particles ($\epsilon \gtrsim 0.9$), however, 
the rise time might differ significantly from less elastic cases.}
\item{The occurrence of the BNE depends on the values of friction constants (see Section~\ref{results_mu}). 
Both too high and too low friction constants diminish convection and thus the BNE.}
\item{The critical condition for the BNE agrees well with the limits of 
$\tilde{\Gamma}\gtrsim1$ and $\tilde{v}_{osc}\gtrsim\tilde{v}_c\sim1$ (see Section~\ref{results_vel}). 
These conditions also agree well with previous simulations and experiments that have comparable densities for small and large particles.}
\item{The BNE is scalable for different gravitational environments by choosing the oscillation frequency 
that corresponds to the gravitational acceleration (i.e., $\omega\propto \sqrt{a_g}$, see Section~\ref{results_lowg}). 
Thus, the same level of size sorting is expected for a smaller oscillation speed on a small asteroid compared to a big one.}
\item{The rise speed of an intruder is proportional to the oscillation speed (see Sections~\ref{results_vel} and \ref{results_lowg}). Also, there might be 
an optimal bed depth for a certain oscillation speed to achieve the fast rise of an intruder.}
\item{For a wide cylinder, the convection roll along the wall may be thick enough to pass the intruder downward, leading 
to a ``whale'' effect (see Section~\ref{results_vel}).}
\item{The BNE is expected to occur on an asteroid for seismic speed that is comparable to non-destructive impacts (see Section~\ref{results_lowg}). 
We have compared the critical oscillation speed for the BNE with the critical seismic speed, and found that 
the region for the BNE overlaps with that of the seismic shaking.}
\item{We also estimate the rise time of a large block from the central region to the surface of an asteroid, and 
predict that multiple impacts or long-lasting seismic shaking might be required for the BNE to significantly change the asteroid surface. A potential 
problem which could affect such multiple impacts is that the BNE might slow down significantly for randomly oriented oscillations.}
\end{enumerate}

The efficiency of the BNE depends on many properties.  In this paper, we have explored 
the dependence on the coefficients of restitution, the friction constants, the oscillation 
frequency and amplitude, the particle bed depth, and the gravity.
One of the potential directions for the future study would be to investigate how the BNE 
changes from the dense limit to the vibro-fluidized regime.  
In particular, it would be interesting to perform the BNE experiments in the high-speed 
region to understand whether the BNE keeps on occurring as our code predicts, or turns off 
sharply as suggested by \shortcite{Godoy2008PhRvE}. 

The rise time of an intruder is important to estimate the efficiency of the BNE on asteroids.
Our work shows that the rise speed linearly scales with the oscillation speed and 
is proportional to $\sqrt{a_g}$. 
Although this result is intuitive, we should check whether such a trend would hold 
for different shaking models as well since most of our simulations assume sinusoidal 
oscillations.   
Interestingly, the analytical model developed by \shortcite{Jiongming1998NCimD} also estimates that the rise 
speed is proportional to $\sqrt{a_g}$.

In this paper, we did not mention the effects of the particle's size distribution, 
the container width, or the container shape.
We assume the idealized system where all of the small particles have the same size 
(except for Section~\ref{results_Tancredi12}) and the size ratio of large to small 
particles is $d_l/d_s=3$.  
However, particles in actual asteroids will generally follow some size distribution.     
Previous studies have shown that the BNE occurs only for a radius ratio of large and small 
particles greater than some threshold \cite[e.g.,][]{Duran1994PhRvE,Vanel1997PhRvL}. 
The size ratio we chose for our simulations ($d_l/d_s=3$) is near the threshold 
according to these experiments.
Thus, we expect that the size ratio might need to be near this value or larger for 
the BNE to take place. 
Also, a preliminary study we performed indicates that the rise speed of an intruder slows down 
when a size distribution of small particles is introduced. 
Such effects on the rise speed should be studied more carefully in future work.

Differences due to the container could also be significant. The experimental studies 
by \shortcite{Hejmady2012PhRvE} suggest that both bed heights and widths affect the rise time 
of an intruder. 
Furthermore, previous studies have shown that ``reverse'' convection (i.e., in which 
the particle flow ascends along the wall and descends around the center) is seen for 
granular materials in a container with outwardly slanting walls 
\cite[e.g.,][]{Grossman1997PhRvE} and thus the intruder is trapped at the bottom rather 
than at the bed surface in such a container \cite[e.g.,][]{Knight1993PhRvL}.
Indeed, when we use a bowl instead of a cylinder as a container, we observe that the intruder goes up 
and down in the granular bed, but never comes to the surface for a density ratio of 
$\rho_l/\rho_s=1$.  Our further tests with $\rho_l/\rho_s=0.5$ and $2$ show that the BNE 
occurs for the former, but not for the latter.  The rise of the intruder in the former 
case is consistent with the expected behavior of the normal fluid. 
There is a related issue when we consider the BNE in an asteroid. 
The BNE requires the container wall to create the convective flow, but there is no 
actual wall such as the ones we considered here in an asteroid.  
However, it is plausible that a very large body could act as a wall for the smaller particles. 
For future work, we intend to investigate the effect of having no lateral walls 
by adopting periodic boundary conditions, and also to improve our current study by 
modelling self-gravitating rubble piles.

One of the most important fundamental problems is the absence of knowledge of the coefficients 
of restitution and friction constants of particles in asteroids.  
Currently, our knowledge of these values relies on experiments, for example, 
the collision experiment done by \shortcite{Durda2011Icar} on 1\,m-size granite spheres or 
the avalanche experiment done by \shortcite{Yu2014inprep} with gravels.
However, it is not clear whether such objects have the same mechanical behavior as 
materials composing asteroids of interest.
There are some future missions that are expected to return asteroid samples, 
such as Hayabusa~2, OSIRIS-REx, and potentially MarcoPolo-R. These efforts will lead 
to a better understanding of surface features and will provide invaluable knowledge of 
asteroid compositions as well as their response to external solicitations, 
such as that of the sampling tool.
\section*{Acknowledgements}
We thank Carsten G\"{u}ttler for useful discussions, and the referee, Gonzalo Tancredi, 
for careful reading and detailed comments.  
SM is supported by an Astronomy Center
for Theory and Computation Prize Fellowship at the University of
Maryland, and also by the Dundee Fellowship at the University of Dundee. 
PM and SRS acknowledge support from the french space agency CNES.
All of the simulations in this paper were done on the YORP cluster in the 
Department of Astronomy at UMD and on the Deepthought 
High-Performance Computing Cluster at UMD. 
This material is based partly on work supported by 
the US National Aeronautics and Space Administration under Grant Nos. NNX08AM39G, 
NNX10AQ01G, and NNX12AG29G issued through the Office of Space Science and by 
the National Science Foundation under Grant No. AST1009579.
\bibliographystyle{mn2e}
\bibliography{REF}
\label{lastpage}
\end{document}